\newif\ifShowKeys
\ShowKeystrue
\ShowKeysfalse

\newif\ifshowtikz
\showtikztrue
\showtikzfalse   % <---- comment/uncomment that line

\documentclass[11pt]{article}
	\pdfoutput=1
\usepackage[T1]{fontenc}

\usepackage{datetime}
\usepackage{comment}					% to comment large parts of text

\usepackage{color}
\definecolor{Maroon}{rgb}{0.68,0.088,0.218}
\definecolor{Mahogany}{rgb}{0.65,0.0975,0.0845}
\usepackage[setpagesize=false,pagebackref=false, 
linktocpage, bookmarksopen=true, colorlinks=true, 
linkcolor=Maroon,citecolor=Maroon,urlcolor=Maroon]{hyperref}

\usepackage[parsep]{collref}				% collect references in groups

\ifShowKeys \usepackage[notcite]{showkeys} \fi

\usepackage{amsmath, amssymb,amsthm}
\usepackage{stackrel}
\numberwithin{equation}{section}
\usepackage{bm,environ,mathrsfs,array,arydshln}
\usepackage{booktabs,float,slashed}
\usepackage{appendix}
\usepackage[mathcal]{euscript}
\usepackage{tensor} 						% Ratcliffe package to write tensors
\usepackage{mathabx}
\usepackage[vcentermath]{youngtab}
\usepackage{simpler-wick}

\usepackage{graphicx,epsfig,epic}
\usepackage{subcaption,wrapfig}
\usepackage{tikz}
\usepackage{tikz-feynman} 
\allowdisplaybreaks

\usepackage{framed}						% for shaded equations \begin{shaded} + \end{shaded} or \bs + \es
\definecolor{shadecolor}{rgb}{0.9996078, 0.984314, 0.960784}
\definecolor{framecolor}{rgb}{0,0,0}
\definecolor{TFTitleColor}{RGB}{1,1,1}
\definecolor{TFFrameColor}{RGB}{249	218	181}		
\definecolor{TFFrameColor}{RGB}{230 230 230 }

\newenvironment{frshaded}{%
    \MakeFramed {\FrameRestore}}%
    {\endMakeFramed}

\definecolor{myred}{RGB}{233, 33, 45}

\usepackage{mdframed}
\definecolor{lightpeach}{RGB}{255, 247, 235}

\newmdenv[
  backgroundcolor=gray!5,
  linecolor=gray!40,
  roundcorner=5pt,
  innerleftmargin=8pt,
  innerrightmargin=8pt,
  innertopmargin=6pt,
  innerbottommargin=6pt,
]{remarkbox}

\newcommand{\bs}{\begin{frshaded}}			% framed with background in shadecolor 
\newcommand{\es}{\end{frshaded}\noindent}

\def\ba#1\ea{\begin{align}#1\end{align}}		        %  clever way to bypass the known problem...
\newcommand{\be}{\begin{equation}}
\newcommand{\ee}{\end{equation}}
\newcommand{\bea}{\begin{equation} \begin{aligned}} 
\newcommand{\eea}{\end{aligned} \end{equation}}
\newcommand{\mc}{\mathcal }
\newcommand{\wh}{\widehat}
\newcommand{\wt}{\widetilde}

\newcommand{\la}{\label}
\newcommand{\eps}{\varepsilon}

\newcommand{\lp}{\notag \\ & }

\DeclareMathOperator{\Tr}{\text{Tr}}

\newcommand{\cf}{\textit{cf.} }
\newcommand{\ie}{\textit{i.e.} }
\newcommand{\eg}{\textit{e.g.} }

\newcommand{\sql}{\sqrt\l}
\renewcommand{\l}{\lambda}

\newcommand{\ket}[1]{|#1\rangle}

\newcommand{\braket}[2]{\langle #1|#2\rangle}
\newcommand{\mmm}[3]{\langle #1|#2|#3\rangle}

\newcommand{\alp}{\alpha}

\newcommand{\eee}{\varepsilon}

\makeatletter
\DeclareFontFamily{OMX}{MnSymbolE}{}
\DeclareSymbolFont{MnLargeSymbols}{OMX}{MnSymbolE}{m}{n}
\SetSymbolFont{MnLargeSymbols}{bold}{OMX}{MnSymbolE}{b}{n}
\DeclareFontShape{OMX}{MnSymbolE}{m}{n}{
    <-6>  MnSymbolE5
   <6-7>  MnSymbolE6
   <7-8>  MnSymbolE7
   <8-9>  MnSymbolE8
   <9-10> MnSymbolE9
  <10-12> MnSymbolE10
  <12->   MnSymbolE12
}{}
\DeclareFontShape{OMX}{MnSymbolE}{b}{n}{
    <-6>  MnSymbolE-Bold5
   <6-7>  MnSymbolE-Bold6
   <7-8>  MnSymbolE-Bold7
   <8-9>  MnSymbolE-Bold8
   <9-10> MnSymbolE-Bold9
  <10-12> MnSymbolE-Bold10
  <12->   MnSymbolE-Bold12
}{}

\let\llangle\@undefined
\let\rrangle\@undefined
\DeclareMathDelimiter{\llangle}{\mathopen}%
                     {MnLargeSymbols}{'164}{MnLargeSymbols}{'164}
\DeclareMathDelimiter{\rrangle}{\mathclose}%
                     {MnLargeSymbols}{'171}{MnLargeSymbols}{'171}
\makeatother

\begin{document}

\begin{comment}
\centerline{\Large\sc  Krylov (DS)SYK notes}
\vskip 0.2cm
\centerline{-- Notes --}
\vskip 0.2cm\centerline{\small\today\ -- \currenttime}
\vskip 0.5cm
 %\centerline{\sc M. Beccaria}
\bigskip
\begin{abstract}
\begin{center}
%\includegraphics[width=0.5\textwidth]{cover}
\end{center}
\end{abstract}
\end{comment}

%%%%%%%%%%%%%%%%%%%%%%%%%%%%%%%%%%%%%%%%%%%%%%%%%%%%%%%%%
%%%%% NOTICE the title page is commented out with \begin{comment}...\end{comment}   but is ready to be used

%\begin{comment}

\begin{titlepage}
%\begin{tabbing}
%\hspace*{11.5cm} \=  \kill % set the tabbings
%\>  Imperial-TP-AT-2024-?? \\
%\> %none
%\end{tabbing}

%\centerline{\small\today\ -- \currenttime}

\vspace*{15mm}
\begin{center}
{\Large  Higher-loop wormhole length in sine-dilaton gravity  \vskip 0.25cm from DSSYK Krylov complexity} %\vskip 9pt
%{\Large\sc      Notes}
\vspace*{10mm}

E. Alfinito$^{a,b}$, M. Beccaria$^{a,b}$

\vspace*{4mm}
{\small
	
${}^a$ Universit\`a del Salento, Dipartimento di Matematica e Fisica \textit{Ennio De Giorgi},
\vskip 0.2cm
${}^{b}$ INFN - sezione di Lecce, Via Arnesano, I-73100 Lecce, Italy
\vskip 0.3cm
\vskip 0.2cm {\small E-mail: \texttt{matteo.beccaria@le.infn.it}}
}
\vspace*{0.8cm}
\end{center}

\begin{abstract}  
The quantum wormhole length in sine-dilaton gravity has been shown to equal the Krylov spread complexity in the double-scaled SYK model.
In the infinite temperature limit, we compute the five-loop  semiclassical expansion of 
DSSYK complexity by singular perturbation of the operator Liouville-type equations of motion,  extending the existing one-loop results.
The same method is applied to evaluate the Krylov variance and third-order cumulant, related to the connected  two- and three-point 
functions of the length operator at coincident points.
The small- and large-time behaviour of these observables is also studied. In particular, for the large-time slope of the wormhole linear growth, 
we determine the all-order resummation of the perturbative series, and the leading non-perturbative corrections.
\end{abstract}
\vskip 0.5cm
	{
		%Keywords: 2D Gravity, AdS-CFT Correspondence, Black Holes, Random Systems
	}
\end{titlepage}

%\end{comment}
%%%%%%%%%%%%%%%%%%%%%%%%%%%%%%%%%%%%%%%%%%%%%%%%%%%%%%%%%

{\small
\makeatletter
\newcommand*{\toccontents}{\@starttoc{toc}}
\makeatother
\toccontents
}

%\tableofcontents

\vspace{1cm}

\setcounter{footnote}{0}

\section{Introduction and summary of results}

The Sachdev-Ye-Kitaev (SYK) model \cite{Sachdev:1992fk,Sachdev:2010um,Maldacena:2016hyu} is a quantum mechanical 
model of $N$  Majorana fermions with random all-to-all $p$-local interactions. 
The model is quantum chaotic, with level repulsion \cite{Cotler:2016fpe} and maximal chaos exponent \cite{Maldacena:2015waa}. 

At large $N$ and fixed $p$, the effective low-energy description of the SYK model
is Schwarzian quantum mechanics, \ie the boundary action of JT gravity on nearly AdS$_{2}$
\cite{Maldacena:2016hyu,Maldacena:2016upp,Jensen:2016pah,Sarosi:2017ykf,Berkooz:2018jqr,Mertens:2017mtv,Engelsoy:2016xyb,Mertens:2022irh}.
While the holographic bulk dual of the complete SYK model is not known, much progress has been made
in the double-scaled model (DSSYK) where the large $N$ limit is taken with fixed $\l = 2p^{2}/N$ 
\cite{Cotler:2016fpe,Berkooz:2018jqr,Berkooz:2018qkz,Berkooz:2024lgq}. 
JT gravity is recovered in the DSSYK model by taking small $\l$ and focusing on the low energy dynamics -- the so-called triple scaled limit
\footnote{
The amplitudes in DSSYK were computed by chord diagrams 
in \cite{Berkooz:2018jqr,Berkooz:2018qkz,Berkooz:2024lgq} and exhibit similarities with 
those in JT gravity \cite{Mertens:2017mtv,Yang:2018gdb,Mertens:2018fds,Blommaert:2018oro,Iliesiu:2019xuh}.}.

At generic $\l$ (and temperature), the  holographic dual to the DSSYK was finally identified with 2d sine-dilaton gravity 
in \cite{Blommaert:2023opb,Blommaert:2024ymv,Blommaert:2024whf} by showing equivalence between the DSSYK transfer matrix \cite{Lin:2022rbf}
and a canonical transformation of the ADM Hamiltonian, adopting 
a suitable choice of operator ordering. Correlation functions can also be matched by comparing chord diagram computations in DSSYK
and gravity results, obtained by introducing a non-minimally coupled probe in the sine-dilaton  theory  dual to bi-local operators in the 
boundary Schwarzian theory \cite{Blommaert:2024ymv}.

As a manifestation of Susskind's
Complexity$=$Volume conjecture (C=V) \cite{Susskind:2014rva,Stanford:2014jda,Brown:2015lvg,Belin:2021bga,Belin:2022xmt},
it was shown in \cite{Rabinovici:2023yex,Xu:2024gfm,Ambrosini:2024sre} that the Krylov complexity \cite{Nandy:2024evd,Baiguera:2025dkc,Rabinovici:2025otw}
of the infinite-temperature thermofield double state \cite{Maldacena:2001kr}
on the boundary of AdS$_{2}$ admits a bulk description in JT gravity as the length of the two-sided wormhole. \footnote{
In \cite{Aguilar-Gutierrez:2026ogo,Aguilar-Gutierrez:2026nmd} an exact correspondence between Krylov complexity and the wormhole length beyond the semiclassical level was
derived for deformations of the 
DSSYK model which change boundary conditions in the bulk. This follows from canonical quantization, in which 
the chord number is isomorphic to the  bulk length operator in 
sine-dilaton and JT gravity.}
In more detail, the Krylov spread complexity $C(t)$ associated with the zero chord state in DSSYK 
was shown in \cite{Rabinovici:2023yex,Ambrosini:2024sre} to be
\be
\la{1.1}
C_{\beta=0}(t) = \frac{2}{\l}\log\cosh(Jt)+O(\l^{0}),
\ee
where $J$ is related to the strength of the SYK random couplings.
In suitable units, and after multiplication by $\l$, this was shown to be the same as the 
length $L_{\rm JT}$ of the wormhole anchored at boundary time $t$ in the classical JT gravity \cite{Harlow:2018tqv,Harlow:2021dfp,Brown:2018bms}
\be
L_{\rm JT} = \lim_{\l\to 0}\l\, C_{\beta=0}(t) = 2\log\cosh(Jt).
\ee
This remarkable result, obtained in the triple scaled limit of DSSYK -- in particular at $\l\to 0$ -- 
was extended to the sine-dilaton gravity context in \cite{Heller:2024ldz}, and further to the de Sitter sector of sine-dilaton gravity 
-- corresponding to the high-energy limit of DSSYK -- in \cite{Heller:2025ddj}.
Starting from the gravity action \cite{Blommaert:2024ymv,Blommaert:2023opb}
\be
\la{1.3}
S = \frac{1}{2}\int_{\mc M} d^{2}\sigma\sqrt{g}\bigg(\Phi R+\frac{2\sin(\l\Phi)}{\l}\bigg)+\text{boundary terms},
\ee
the gravitational two-point function corresponds to  the boundary-to-boundary propagator of a non-minimally coupled 
scalar field in the bulk \cite{Blommaert:2024ymv}. After a rescaling of the dilaton, the coupling $\l$ is a semiclassical expansion parameter
and the classical geometry for $\l\to 0$ is an AdS$_{2}$ black hole with Hawking temperature parametrized as $\beta_{\rm BH} = 2\pi/\sin\theta$. 
After holographic renormalization, the wormhole length reads
\be
\la{1.4}
L = 2\log\cosh\frac{t\sin\theta}{2}-2\log\sin\theta+O(\l),
\ee
in terms of the two-sided boundary Lorentzian time $t$.
This is the semiclassical limit of the full quantum expectation value of the  wormhole length operator $\hat L$
in sine-dilaton gravity \cite{Blommaert:2024ymv}, evaluated on the thermofield double state \cite{Maldacena:2001kr},
as a function of time. As discussed in \cite{Blommaert:2024whf}, the relation between $\theta$ and the microscopic DSSYK inverse temperature $\beta$
is  given by
\be
\la{1.5}
\beta = \frac{2\pi-4\theta}{\sin\theta},
\ee
and, in particular, the limit $\beta\to 0$ corresponds  to $\theta\to \pi/2$. \footnote{For further discussion of the so-called \textit{fake} 
periodicity $\beta_{\rm BH}$, see \cite{Lin:2023trc}.}
In sine-dilaton gravity, the two-point function of a massive probe is captured by the insertion of the bilocal operator $e^{-\Delta\hat L}$ 
of conformal weight $\Delta$ \cite{Blommaert:2024ymv}. Following \cite{Iliesiu:2021ari}, in \cite{Heller:2024ldz} the length expectation value is extracted from 
\be
\la{1.6}
\langle\hat L\rangle = \left[-\partial_{\Delta}\langle e^{-\Delta\hat L}\rangle\right]_{\Delta=0},
\ee
where $\langle e^{-\Delta\hat L}\rangle$ is computed (at leading topological order) by dividing the disk 
boundary into two parts of length $\tau$ and $\beta-\tau$ corresponding to the Euclidean boundary times to which the semiclassical geodesic is anchored.
%
%After analytic continuation to Lorentzian time 
%\be
%\la{1.7}
%\tau = \frac{\beta}{2}+it, 
%\ee
The key relation proved in  \cite{Heller:2024ldz} is 
\be
\la{1.7}
\langle\hat L\rangle = \l\, C_{\beta}(t),
\ee
where Lorentzian time $t$ is related to $\tau$ by $\tau=\frac{\beta}{2}+it$, and $C_{\beta}(t)$ is the DSSYK Krylov complexity evaluated after 
mixed Lorentzian ($t$) and Euclidean ($\beta/2$) evolution of the zero-chord state.

Notice that use of  (\ref{1.6}), \ie the relation to the 2-point function of a suitable conformal operator, is convenient in sine-dilaton theory and 
optional in DSSYK. For instance,  such a relation was not exploited in the  derivation of (\ref{1.1})  \cite{Rabinovici:2023yex}. When relation (\ref{1.6})
is used in DSSYK context, the operator $e^{-\Delta \hat L}$ corresponds to a string of fermions, and its expectation value 
is known in closed form, see \eg \cite{Berkooz:2018jqr,Goel:2023svz}, although its semiclassical expansion in small $\l$ is non-trivial. 
At one-loop order, it has been worked out in  \cite{Goel:2023svz,Okuyama:2023bch}. The classical contribution is 
\be
\la{1.8}
\langle e^{-\Delta \hat L}\rangle_{\rm cl} = \bigg(\frac{\sin^{2}\theta}{\sin^{2}(\sin\theta\, \frac{\tau}{2}+\theta)}\bigg)^{\Delta}.
\ee
Using (\ref{1.5}) and (\ref{1.6}), we reproduce (\ref{1.4}) and, 
%\be
%\la{1.9}
%\langle \hat L\rangle_{\rm cl} = 2\log\cosh\frac{t\sin\theta}{2}-2\log\sin\theta,
%\ee
%which agrees with (\ref{1.4}). 
in the infinite temperature limit $\beta=0$ (or $\theta=\frac{\pi}{2}$) we specialize to (\ref{1.1}), where $J=1/2$ is implicit in the sine-dilaton/DSSYK literature.
%\be
%\la{1.11}
%\langle \hat L\rangle_{{\rm cl}, \beta=0} = 2\log\cosh\frac{t}{2},
%\ee
%that agrees with (\ref{1.1}). 
%As a remark, we notice that (\ref{1.1}) was obtained by direct semiclassical expansion of Krylov complexity, \ie without resorting to the relation 
%with the bilocal operator in (\ref{1.6}). }
At one-loop order, the saddle point analysis of $\langle e^{-\Delta\hat L}\rangle$ can be carried out by expanding in small $\Delta$. At quadratic order, it 
reads \cite{Goel:2023svz,Okuyama:2023bch} 
\be
\la{1.9}
\langle e^{-\Delta\hat L}\rangle = \langle e^{-\Delta\hat L}\rangle_{\rm cl}\,\bigg[1+\frac{1}{2}\l\,\bigg(\Delta\mc A_{1}+\Delta^{2}\mc A_{2}+O(\Delta^{3})\bigg)+O(\l^{2})\bigg],
\ee
where the first two one-loop coefficients are 
\bea
\la{1.10}
\mc A_{1} &= \frac{1}{2(1+u \tan u)}\bigg[-\frac{(1+u \tan u)^{2}}{\cos^{2}\zeta}+\frac{(1+\zeta\tan \zeta)^{2}}{\cos^{2}u}+\zeta^{2}(\tan^{2}u-\tan^{2}\zeta)
-\frac{1+\zeta\tan\zeta}{1+u \tan u}+1\bigg], \\
\mc A_{2} &= -\frac{(\tan\zeta+\tan\zeta(\zeta+u)\tan u+\tan u)(\tan\zeta+\tan u(\tan\zeta(u-\zeta)-1))}{1+u\tan u},
\eea
with
\be
\la{1.11}
u = \frac{\pi}{2}-\theta, \qquad \zeta = \frac{\pi}{2}-\theta-\frac{\tau}{2}\sin\theta.
\ee
In \cite{Bossi:2024ffa}, the one-loop result (\ref{1.9}) was reproduced in  the quantized sine-dilaton theory \footnote{More precisely, they considered a 
description of sine-dilaton gravity in terms of the $q$-Schwarzian theory, a deformed version of the standard Schwarzian model dual
to JT gravity. A recent analysis of JT gravity and general dilaton gravities at finite cutoff, using both the bulk path integral and the path integral over boundary curves, can be found in \cite{Griguolo:2025kpi}.
}
by a careful use of Gelfand-Yaglom methods to deal with the relevant one-loop functional determinants. In the infinite temperature limit, 
the above expressions simplify to 
\ba
\la{1.12}
\mc A_{1} &= \frac{1}{4}\tau\tan\frac{\tau}{2}-\frac{1}{2}\tan^{2}\frac{\tau}{2}, \qquad
\mc A_{2} =-\tan^{2}\frac{\tau}{2},
\ea
and we get the 1-loop correction to (\ref{1.1})
\be
\la{1.13}
C_{\beta=0}(t) = \frac{2}{\l}\log\cosh\frac{t}{2}+\frac{1}{8}\,t\,\tanh \frac{t}{2}-\frac{1}{4}\tanh^{2}\frac{t}{2}+O(\l).
\ee

\paragraph{Summary of results}

In principle, the 1-loop correction in (\ref{1.13}) as well as similar higher order contributions could be reproduced by applying 
the methods  used in \cite{Rabinovici:2023yex} to obtain the leading order result (\ref{1.1}).
Alternatively, relying on (\ref{1.6}), one could try to get (\ref{1.13}) by extension of the saddle-point analysis in \cite{Goel:2023svz,Okuyama:2023bch},
and taking $\beta=0$ as a special limit. As we will discuss, neither of these two approaches works in a straightforward way.

For this reason, we  present here a different approach to the semiclassical expansion of the DSSYK spread complexity that 
can be extended to higher orders with moderate effort. The method is based on the singular $\l\to 0$ expansion of the operator equations of motion 
of the $\hat L$ operator in the DSSYK model which take the form of a one-dimensional Liouville equation. 

Our main result is the 5-loop extension of (\ref{1.13}). It reads
\be
\la{1.14}
C_{\beta=0}(t) = \frac{1}{\l}\langle \hat L \rangle = K_{0}(Jt)\frac{1}{\l}+K_{1}(Jt)+K_{2}(Jt)\,\l+K_{3}(Jt)\,\l^{2}+K_{4}(Jt)\,\l^{3}+K_{5}(Jt)\,\l^{4}+\cdots,
\ee
where the new functions $K_{2}, \dots, K_{5}$ are explicit polynomials in $\tanh x$ with $x$-dependent coefficients (as in the 1-loop correction). Their explicit expression is 
collected in (\ref{1.15}) in Section~\ref{sec:results}.
By the same approach, it is also possible to obtain loop expansions for higher-order Krylov complexities or cumulants of $\hat L$. These observables
have been considered in the complexity literature in 
\cite{Bhattacharjee:2022lzy,Fu:2024fdm,Camargo:2024rrj} and in the holographic context in 
\cite{Iliesiu:2021ari,Bhattacharjee:2022ave,Fu:2025kkh,Grabarits:2025xys,Alfinito:2026vah}, and provide a measure of quantum fluctuations
of the length operator.
By applying our method, we find the following small $\l$ expansion \footnote{We recall that the $p$-th cumulant is defined by 
$\langle \mc O^{p}\rangle_{\rm conn} = \frac{d^{p}}{dz^{p}}\log\langle e^{z \mc O}\rangle|_{z=0}$. The normalization in (\ref{1.17})
is based on the identification between $\hat L/\l$ and the Krylov number operator.
}
\ba
\la{1.17}
C_{\beta=0}^{(p)}(t) \equiv \frac{1}{\l^{p}}\langle \hat L^{p}\rangle_{\rm conn} &= K^{(p)}_{0}(Jt) \frac{1}{\l}+K^{(p)}_{1}(Jt)+K^{(p)}_{2}(Jt)\, \l+\cdots, 
\ea
where for the Krylov variance ($p=2$) the coefficients up to 4-loop order, \ie $K_{0}^{(2)}, \dots, K^{(2)}_{4}$  are given in (\ref{1.18})  in Section~\ref{sec:results}.
For the third-order cumulant ($p=3$)  we computed the expansion up to 3-loop order, \ie $K^{(3)}_{0}, \dots, K^{(3)}_{3}$ that are collected in (\ref{1.19}) in the same section.

As an application of the formalism and the results, we examine the 
small and large time asymptotic behaviour of the Krylov complexity and of the higher cumulants. At small time, we prove that 
\be
\la{116}
C^{(p)}(t) = \frac{(Jt)^{2}}{\l}-\frac{1}{12}(3\cdot 2^{p}-4)(1-e^{-\l})\,\frac{(Jt)^{4}}{\l^{2}}+\cdots,
\ee
where higher corrections in $Jt$ are easily computable in our approach. \footnote{Note that (\ref{116}) is  an exact-in-$\l$ 
expansion in powers of $(Jt)^{2}$.}
The determination of the large-time behaviour is more difficult.
The functions $K^{(p)}_{n}(Jt)$ are polynomials in $\tanh(Jt)$ with coefficients that are polynomials in $Jt$, as is the case in the 1-loop correction
in (\ref{1.13}). At large time, we can approximate $\tanh(Jt)\simeq 1$ up to exponentially small corrections. In this limit, the Krylov complexity 
takes the form \footnote{The suppressed terms carry enhancement factors that are powers of $t$, as mentioned above.}
\be
C(t) = A_{0}(\l)+A_{1}(\l)\,Jt+O(e^{-Jt}),
\ee
\ie it is linear in time with $\l$-dependent coefficients. All higher powers of $t$  cancel at each loop order.
The small $\l$ expansion of the coefficients $A_{0},A_{1}$ is extracted from the 5-loop expansion of the 
Krylov complexity and reads
\be
\la{117}
A_{0}(\l) = -\frac{2\log 2}{\l}-\frac{1}{4}-\frac{7\l}{96}-\frac{\l^{2}}{64}-\frac{211\l^{3}}{46080}-\frac{7\l^{4}}{3072}+\cdots,\ \
A_{1}(\l) = \frac{2}{\l}+\frac{1}{4}+\frac{7\l}{192}+\frac{5\l^{2}}{1536}-\frac{\l^{3}}{81920}-\frac{101\l^{4}}{5898240}+\cdots.
\ee
A discussion of (\ref{116}) and (\ref{117}) in comparison with existing estimates is presented. 
%Although our results are analytically derived and thus exact, 
%we also provide supporting numerical checks at moderate $\l$, where the role of subleading terms becomes visible. 
The expansions (\ref{117}) are valid in the small $\l$ regime. 
A relatively simple conjecture for the resummed form 
of the terms in the slope $A_{1}(\l)$ is 
\be
\la{119}
A_{1}^{\rm pert}(\l) = \frac{2}{\l}e^{\l/8}\sqrt{\frac{\l/4}{\tanh(\l/4)}} = \frac{2}{\lambda }+\frac{1}{4}+\frac{7 \lambda }{192}
+\frac{5 \lambda^2}{1536}-\frac{\lambda ^3}{81920}-\frac{101 \lambda^4}{5898240}
+\frac{2473 \lambda ^5}{1981808640}+\cdots,
\ee
which matches the first six terms of the small $\l$ expansion from our 5-loop calculation. 
We prove this formula analytically and also give the leading non-perturbative correction:
\be
A_{1}^{\rm non-pert}(\l) = -\frac{2\sqrt 2}{\pi^{7/2}}\l^{3/2}e^{-\frac{\pi^{2}}{2\l}}+\cdots.
\ee
Higher order non-perturbative corrections can be extracted from the formulas in the main text.

The analysis of the higher cumulants is similar and it turns out that they are polynomials in $t$ of degree $p$ -- up to 
exponentially suppressed terms -- again with $\l$-dependent coefficients,
as suggested by the $p=2,3$ cases. In this case, we also provide the analytical form of the resummed perturbative part of the leading large-time
coefficient $A_{p}^{(p)}(\l)$, which is given in  (\ref{res-cum}).

\paragraph{Final remarks} In our analysis, we stopped at 5-loop order, but the proposed method is algorithmic and 
can be pushed beyond this order with more computational effort, if needed. In our opinion, the 5-loop expansion is already a 
non-trivial extension of the known 1-loop result. Of course, alternative approaches exist, and we discuss in some detail their possible
limitations when pushed at higher loop order. 

The major limitation of the proposed method is that it is tailored to the infinite temperature limit $\beta=0$.
As discussed in \cite{Heller:2024ldz}, at non-zero $\beta$
one has to replace $t\to t-i\beta/2$
in the Krylov amplitude entering the calculation of Krylov complexity. Since complexity involves the squared modulus of the amplitude, we get 
two factors with shifts $t \pm i\beta/2$
in opposite directions in the complex plane, and this is not a trivial analytic continuation of the Krylov complexity evaluated at real times. 
Nevertheless, within our framework,  the finite-$\beta$
problem reduces to a modification of the initial conditions of the equations of motion, while their structure is unchanged. This suggests that a perturbative expansion in $\beta$
around the infinite temperature limit is feasible, and we leave this to future work.

A further physical limitation of the semiclassical expansion that should be kept in mind is that it captures only the initial growth regime of the complexity, 
where the chord basis is a good approximation to the physical Krylov basis. The non-perturbative saturation of the wormhole size at late times, 
associated with the finite dimensionality of the full Hilbert space \cite{Nandy:2024zcd,Balasubramanian:2024lqk}, 
lies beyond the reach of the perturbative $\lambda$ expansion in full chord space.

\paragraph{Plan of the paper}

In Section~\ref{sec:chord}, we recall the structure of the DSSYK Hamiltonian in chord space, and in Section~\ref{sec:krylov} we define Krylov complexity and discuss its basic properties in the DSSYK setting. 
In Section~\ref{sec:overview}, we examine the limits of two existing approaches to the semiclassical expansion: direct saddle-point analysis of the Krylov complexity, 
and perturbation theory for the matter operator 2-point function in the $G\Sigma$
formalism. 
In Section~\ref{sec:liouville}, we present the operator equations of motion for the relevant $q$-algebra DSSYK operators in Liouville form. 
These equations are solved as high-order Taylor expansions at small time in Section~\ref{sec:taylor}, providing a useful benchmark for their analytic closed-form resummation, valid at all times, 
which is derived in Section~\ref{sec:singular}. In Section~\ref{sec:results}, we present the 5-loop 
derivation of the Krylov complexity as well as 
loop expansions for the Krylov variance at 4-loops 
and the third-order cumulant at 3-loop order. Finally, in Section~\ref{sec:asym}, we discuss the small and large time behaviour of the Krylov complexity
and cumulants.

\section{The DSSYK model and chord space}
\la{sec:chord}

The Majorana SYK model has $N$ Majorana fermions $\psi_{i}$, $\{\psi_{i}, \psi_{j}\}=2\delta_{ij}$, $i,j=1, \dots, N$ with 
\be
H_{\rm SYK} = i^{p/2}\sum_{1\le i_{1}<i_{2}<\cdots<i_{p}\le N} J_{i_{1}\cdots i_{p}}\ \psi_{i_{1}}\cdots \psi_{i_{p}}\equiv i^{p/2}\sum_{I}J_{I}\Psi_{I},
\ee
where $I$ is an ordered set of $p$ indices and $J_{I}$ are Gaussian random couplings with 
\be
\llangle J_{I}\rrangle_{J}=0, \qquad
\llangle J_{I}J_{K}\rrangle_{J} =\frac{N}{2p^{2}}\binom{N}{p}^{-1}J^{2}\delta_{IK},
\ee
where  $\llangle\bullet\rrangle$ denotes ensemble averaging and 
we adopt the normalization in \cite{Maldacena:2016hyu,Lin:2022rbf,Lin:2023trc}.
The double-scaling (DS)  limit is \cite{Erdos:2014zgc,Cotler:2016fpe}
\be
N,p\to\infty, \qquad \l\equiv\frac{2p^{2}}{N}=\text{fixed}, \qquad q\equiv e^{-\l}.
\ee
In this limit, it is possible to define an effective Hilbert space of DSSYK \cite{Berkooz:2018jqr,Lin:2022rbf} which is spanned by 
orthonormal chord states $\{\ket{\ell}\}_{\ell=0,1,2,\dots}$. In this basis, one introduces the Hamiltonian
\be
\la{2.4}
H = \frac{J}{\sql}(\alp+\bar\alp), 
\ee
where $\alp$ and $\bar\alp$ are $q$-oscillators that together with the chord number $\hat{\ell}$ defined by $\hat\ell\ket{\ell} = \ell\ket{\ell}$,  \footnote{
We use a caret to distinguish the operator $\hat\ell$ from its  eigenvalue. This will not be necessary for other operators; for those we omit the caret to
keep the notation simpler, at the price of some asymmetry.
}
obey 
\cite{Lin:2023trc}
\be
\la{2.5}
[\alpha, \bar\alpha]_{q} = \alpha\bar\alpha-q\,\bar\alpha\alpha=1, \qquad [\hat\ell, \bar\alpha]=\bar\alpha, \qquad
[\hat\ell, \alpha] = -\alpha,
\ee
and one proves that 
\be
\llangle\Tr f(H_{\rm SYK})\rrangle \stackrel{DS}{\to} \mmm{0}{f(H)}{0}.
\ee
From the holographic point of view, the 0-chord state is dual to the infinite temperature thermofield double entangled state in the gravitational theory
\cite{Lin:2022rbf}.

The algebra (\ref{2.5}) is realized in chord space by 
\be
\la{2.7}
\hat\ell\ket{\ell} = \ell\ket{\ell}, \qquad \bar\alpha\ket{\ell} = [\ell+1]_{q}^{1/2}\ket{\ell+1}, \qquad \alpha\ket{\ell} = [\ell]_{q}^{1/2}\ket{\ell-1},
\ee
where the $q$-integer $[\ell]_{q}$ is 
\be
[\ell]_{q} = \frac{1-q^{\ell}}{1-q}.
\ee
From the explicit realization in (\ref{2.7}), one has the useful operator  relation 
\be
\la{2.9}
\bar\alpha\alpha = [\hat \ell]_{q} = \frac{1-q^{\hat\ell}}{1-q}, 
\ee
as well as the trivial shift identities
\be
\la{2.10}
f(\hat\ell)\,\alpha = \alpha\, f(\hat\ell-1), \qquad f(\hat\ell)\,\bar\alpha = \bar\alpha\,f(\hat\ell+1).
\ee
In the chord basis, the DSSYK Hamiltonian has a tridiagonal form 
\ba
\la{2.11}
H\ket{\ell} &= b_{\ell}\ket{\ell-1} + b_{\ell+1}\ket{\ell+1}, \qquad
b_{\ell} = \frac{J}{\sql}[\ell]_{q}^{1/2}.
\ea
The spectrum of $H$ is known \cite{Berkooz:2018jqr,Berkooz:2018qkz}. A generic eigenstate with energy $E$ obeys
\be
E\psi_{\ell}(E) = \frac{J}{\sqrt{\l(1-q)}}\bigg[\sqrt{1-q^{\ell+1}}\ \psi_{\ell+1}(E)+\sqrt{1-q^{\ell}}\ \psi_{\ell-1}(E)\bigg].
\ee
The spectrum is continuous, with
\be
\la{2.13}
E(\theta) = \frac{2J}{\sqrt{\l(1-q)}}\, \cos\theta, \qquad \theta\in[0,\pi],
\ee
and the normalized eigenvectors are 
\be
\la{2.14}
\psi_{\ell}(\cos\theta) = \sqrt{(q;q)_{\infty}}\, |(e^{2i\theta}; q)_{\infty}|\, \frac{H_{\ell}(\cos\theta|q)}{\sqrt{2\pi(q;q)_{\ell}}},
\ee
where (standard) notation is summarized in Appendix \ref{app:q-functions}.

\subsection{Partition function and 2-point function of matter operators}

The Euclidean partition function in chord space is \cite{Berkooz:2018jqr,Berkooz:2018qkz}
\be
Z(\beta) = \mmm{0}{e^{-\beta H}}{0} = \int_{0}^{\pi}d\theta\, \rho(\theta)e^{-\beta E(\theta)}, \qquad \rho(\theta) = (q, e^{\pm 2i\theta}; q)_{\infty}.
\ee
Matter operators $\mc O_{\Delta}$ with $0<\Delta<1$ may be introduced in DSSYK, and are composites of $p\Delta$  fermions before the 
double-scaling limit, coupled by (new) random
Gaussian couplings \cite{Berkooz:2018jqr}. Their Euclidean 2-point function is 
\be
\la{2.16}
\mmm{0}{\mc O_{\Delta}(\beta_{1})\mc O_{\Delta}(\beta_{2})}{0} = \int_{0}^{\pi}\prod_{i=1,2}d\theta_{i}\, \rho(\theta_{i})e^{-\beta_{i}E(\theta_{i})}
\frac{(q^{2\Delta}; q)_{\infty}}
{(q^{\Delta}, e^{i(\pm \theta_{1}\pm \theta_{2})}; q)_{\infty}}.
\ee

\subsection{Krylov complexity in the DSSYK model}
\la{sec:krylov}

Krylov spread complexity is a versatile notion of quantum information spreading
under Hamiltonian evolution   \cite{Nandy:2024evd,Baiguera:2025dkc,Rabinovici:2025otw}. 
It was introduced in \cite{Parker:2018yvk} as a  probe of chaos finer than 
out-of-time-order correlators, see for instance \cite{Maldacena:2015waa}. Its definition is 
based on the Lanczos algorithm that builds a sequence of states obtained by iteratively applying the 
Hamiltonian to an initial state and performing Gram-Schmidt orthonormalization.

In more detail, we start from a normalized quantum state $\ket{K_{0}}$ and the Hamiltonian operator $H$, and define 
the  states $\{\ket{K_{\ell}}\}_{\ell\ge 1}$ and so-called Lanczos coefficients $\{a_{\ell}\}_{\ell\ge 0}$, $\{b_{\ell}\}_{\ell\ge 1}$ by iterating 
the relations 
\bea
& \ket{A_{\ell+1}} = (H-a_{\ell})\ket{K_{\ell}}-b_{\ell}\ket{K_{\ell-1}}, \qquad \ket{K_{\ell}} = b_{\ell}^{-1}\ket{A_{\ell}},\\
& a_{\ell} =\mmm{K_{\ell}}{H}{K_{\ell}}, \qquad b_{\ell}= \Vert \ket{A_{\ell}}\Vert.
\eea
The states $\{\ket{K_{\ell}}\}$ are orthonormal by construction and may be used as a basis with tridiagonal form of $H$
(we define for convenience $b_{0} \equiv 0$)
\be
\la{2.18}
H\ket{K_{\ell}} = a_{\ell}\ket{K_{\ell}}+b_{\ell}\ket{K_{\ell-1}}+b_{\ell+1}\ket{K_{\ell+1}}.
\ee
The Krylov (spread) complexity of the initial state $\ket{K_{0}}$ is defined as the following function of time 
\be
\la{2.19}
C(t) = \sum_{\ell=1}^{\infty}\ell\, |\braket{K_{\ell}}{K_{0}(t)}|^{2},  \qquad \ket{K_{0}(t)} = e^{-iHt}\ket{K_{0}}.
\ee
Expanding the evolved initial state $\ket{K_{0}(t)}$ in the Krylov basis
\be
\ket{K_{0}(t)} = \sum_{\ell=0}^{\infty}\phi_{\ell}(t)\ket{K_{\ell}},
\ee
the coefficients $\phi_{\ell}(t)$ obey 
\be
\la{2.21}
i\dot\phi_{\ell} = a_{\ell}\phi_{\ell}+b_{\ell+1}\phi_{\ell+1}+b_{\ell}\phi_{\ell-1},
\ee
with boundary condition $\phi_{\ell}(0) = \delta_{\ell,0}$. For the DSSYK Hamiltonian, the chord states are 
the Krylov basis associated with the initial 0-chord state $\ket{0}$, as follows from comparing 
(\ref{2.18}) and  (\ref{2.11}), \ie $\ket{K_{\ell}} = \ket{\ell}$.  In particular, the Lanczos coefficients read
\be
\la{2.22}
a_{\ell}=0, \qquad b_{\ell}= \frac{J}{\sql} [\ell]_{q}^{1/2}.
\ee
Using the known spectrum and eigenfunctions of $H$, \cf (\ref{2.13}), (\ref{2.14}), one obtains the exact expression \cite{Rabinovici:2023yex}  
\ba
\la{2.23}
C(t) &= \sum_{\ell=1}^{\infty}\ell\,|\phi_{\ell}(t)|^{2} = (q;q)_{\infty}^{2}\int_{0}^{\pi}\frac{d\theta}{2\pi}\int_{0}^{\pi}\frac{d\theta'}{2\pi}\ e^{-\frac{2iJt}{\sqrt{\l(1-q)}}(\cos\theta-\cos\theta')}
|(e^{2i\theta};q)_{\infty}|^{2}|(e^{2i\theta'};q)_{\infty}|^{2}\lp
\qquad\qquad \sum_{\ell=1}^{\infty}\frac{\ell}{(q;q)_{\ell}}H_{\ell}(\cos\theta|q)\, H_{\ell}(\cos\theta'|q).
\ea
This formula  is hard to analyze, but from the structure of Lanczos $b_{\ell}$ coefficients it is argued in \cite{Rabinovici:2023yex} that 
the asymptotic form of $C(t)$ at small or large times is
\be
\la{2.24}
C(t) = \begin{cases}
\frac{(Jt)^{2}}{1-q}, & t\ll t_{*}(\l), \\
\frac{2Jt}{\sqrt{\l(1-q)}}, & t\gg t_{*}(\l).
\end{cases}\qquad 
t_{*}(\l) = \frac{1}{J}\sqrt\frac{1-e^{-\l}}{\l},
\ee
where we note that $t_{*}(\l)\to 1/J$ for $\l\to 0$. By taking a suitable continuum limit of the Krylov chain, 
it is possible to derive an interpolation formula valid at all times in the small $\l$ regime \cite{Rabinovici:2023yex}, 
\be
\la{2.25}
C(t)  = \frac{2}{\l}\log\cosh(Jt)+O(\l^{0}),
\ee
which is the result (\ref{1.1}) mentioned in the Introduction.
We remark, however, that the extension of the continuum limit approach to higher orders in $\l$ is non-trivial.

\section{Overview of existing methods for semiclassical expansion}
\la{sec:overview}

Before presenting our approach to the systematic semiclassical expansion of Krylov complexity in chord space, let us 
discuss existing  methods that have been exploited in the literature. 
The aim of this section is not primarily pedagogical, but rather to examine whether the approaches 
used previously can be systematically extended to higher loops, and to identify the obstacles that arise in doing so.

\subsection{Direct saddle point expansion of Krylov complexity}

Since we have an exact explicit expression of the Krylov complexity in (\ref{2.23}), it is natural to examine its direct 
expansion in the $\l\to 0$ limit, or $q\to 1$. This approach is discussed in \cite{Rabinovici:2023yex} and we review it with the aim of considering
its problems. The formula (\ref{2.23}) is equivalent to 
\ba
\la{3.1}
C(t) &= \sum_{n=1}^{\infty}n\,|\phi_{n}(t)|^{2}, \qquad
\phi_{n}(t) = \int_{0}^{\pi}\frac{d\theta}{2\pi}e^{-2iJt\,\frac{\cos\theta}{\sqrt{\l(1-q)}}}\frac{(q; q)_{\infty}}{\sqrt{(q;q)_{n}}}\,|(e^{2i\theta};q)_{\infty}|^{2}\, H_{n}(\cos\theta|q).
\ea
At leading order in small $\l$, one has \cite{Rabinovici:2023yex,Okuyama:2023bch}
\ba
\la{3.10}
\phi_{n}(t) &= \sqrt\frac{2\pi}{\l}\frac{1}{\sqrt{n!}}\l^{-n/2}\lp
\int_{0}^{\pi}\frac{d\theta}{2\pi}(2\cos\theta)^{n}\exp\bigg[-\frac{2iJt}{\l}\cos\theta
-\frac{2}{\l}\bigg(\theta-\frac{\pi}{2}\bigg)^{2}-\frac{iJt}{2}\cos\theta+\log(2\sin\theta)
+O(\l)\bigg].
\ea
The leading order saddle point equation for the saddle position $\theta_{0}$ is \footnote{
One may include the prefactor $(2\cos\theta)^{n}$ in the saddle equation, but this is a correction that does not alter the leading $\l\to 0$ limit.}
\be
\la{3.11}
\pi-2\theta_{0}+i J t \sin\theta_{0}=0.
\ee
Using the parametrization $\theta_{0}=\frac{\pi}{2}+i\gamma$,
the saddle point equation (\ref{3.11}) reads
\be
\la{3.13}
Jt = \frac{2\gamma}{\cosh\gamma}.
\ee
For moderate $Jt \lesssim 1.3$ we get a real solution for $\gamma$ that implies a complex $\theta_{0}$. This is already a complication 
in the sense that one has to control the effect of moving the integration contour in (\ref{3.10}). Setting this issue aside, we 
focus on the small $Jt$ regime where we may solve (\ref{3.13}) perturbatively
\be
\la{3.14}
\gamma = \frac{Jt}{2}+\frac{1}{16}(Jt)^{3}+\frac{13}{768}(Jt)^{5}+\cdots, 
\ee
Changing variable from $\theta$ to $\xi$ according to  
$\theta = \frac{\pi}{2}+i\gamma+\xi\, \sql$,
we get 
\ba
\phi_{n}^{\rm LO}(t) &= \sqrt\frac{2\pi}{\l}\frac{1}{\sqrt{n!}}\l^{-n/2}\, e^{\frac{2\gamma^{2}-2Jt \sinh\gamma}{\l}}e^{-\frac{iJt}{2}}
2\cosh\gamma(-2i\sinh\gamma)^{n}\sql\int_{0}^{\pi}\frac{d\xi}{2\pi}e^{-\xi^{2}(2-Jt \sinh\gamma)}.
\ea
For small enough $Jt$ we can integrate and get 
\be
|\phi_{n}^{\rm LO}(t)|^{2} =  \frac{1+\cosh (2\gamma)}{2-Jt \sinh\gamma}e^{\frac{4}{\l}(\gamma^{2}-Jt \sinh\gamma)}\frac{1}{n!}\bigg(\frac{4\sinh^{2}\gamma}{\l}\bigg)^{n}.
\ee
Substituting this expression in the infinite sum giving the Krylov complexity in (\ref{3.1}), we get
\be
C^{\rm LO}(t) = \frac{1}{\l}\frac{2\sinh^{2}(2\gamma)}{2-Jt \sinh\gamma}e^{\frac{4}{\l}(\gamma^{2}-Jt \sinh\gamma+\sinh^{2}\gamma)}.
\ee
In \cite{Rabinovici:2023yex}, this expression was considered in the limit $Jt\to 0$, although in principle this is in tension with the saddle point treatment because $Jt/\l$ is not necessarily large.
%Notice also that, as we mentioned, higher-order corrections in $\l$ contain terms with $1/\cos^{2}\theta$ that are singular in the small $Jt$ limit.
Substituting the perturbative expansion of the  saddle point (\ref{3.14}),
one obtains 
\be
C^{\rm LO}(t) =  \frac{J^{2}t^{2}}{\l}+O(t^{4}),
\ee
that reproduces the small time behaviour in (\ref{2.24}).
However, the next terms in the small $Jt$ expansion include contributions $\sim t^{6}/\l^{2}$ and so on with higher powers of $1/\l$. These arise from the 
exponential factor
\be
\exp\bigg[\frac{4}{\l}(\gamma^{2}-Jt \sinh\gamma+\sinh^{2}\gamma)\bigg]= \exp\bigg[\frac{1}{4\l}(Jt)^{4}+\frac{1}{9\l}(Jt)^{6}+\cdots\bigg].
\ee
Thus, even recovering the leading order result at small $\l$ in (\ref{2.25}) is non-trivial in this approach. The reason  
is related to a non-uniformity in the Krylov index $n$ and the naive assumption that $(2\cos\theta)^{n}$ is $O(\l^{0})$. 
This could be spoiled if the relevant contribution comes from $n\sim 1/\l$. Actually, this scenario is natural, taking into account that the
$2\log\cosh(Jt)$ result in (\ref{2.25}) comes
from a continuum limit analysis of the Lanczos evolution equation with $\l n = O(\l^{0})$ playing the role of a continuous position on the Krylov chain.

\subsection{Perturbation theory in $G\Sigma$ formalism}

The origin of the difficulties discussed in the previous section is the infinite sum over the Krylov index in (\ref{3.1}). This may in principle be bypassed by 
using  (\ref{1.6}), \ie the relation between complexity and the 2-point function of the DSSYK matter operator with dimension $\Delta$. This approach 
has been used in \cite{Okuyama:2023bch}, see also \cite{Goel:2023svz}, first by a saddle point analysis of the expression (\ref{2.16}). The saddle point equations 
can be treated in an expansion in small $\Delta$, something which is consistent with (\ref{1.6}). The 1-loop correction is computed and, as usual, it  requires 
evaluating the one-loop determinant of fluctuations. Extension to higher loops is in principle feasible.

A more systematic approach, presented in  \cite{Okuyama:2023bch}, which we examine here in some detail, is based on 
perturbation theory in the so-called
$G\Sigma$ formalism where one introduces bi-local antisymmetric fields $G(\tau_{1},\tau_{2})$ and $\Sigma(\tau_{1},\tau_{2})$ 
as in Hubbard–Stratonovich transformation to integrate out the SYK model 
fermions \cite{Sachdev:1992fk}. In the double-scaling limit, $\Sigma$ is integrated out and $G$ is written in terms of a symmetric field $g(\tau_{1}, \tau_{2})$, see for instance \cite{Goel:2023svz}.
If  $0<\tau_{2}<\tau_{1}<\beta$ are ordered points on $S_{\beta}^{1}$, and $\tau_{\pm}=\tau_{1}\pm \tau_{2}$, the action is a Liouville-type and reads
\be
S = \frac{1}{8\l}\int d\tau_{-} d\tau_{+}\ \bigg[-\frac{1}{2}(\partial_{-}g)^{2}+\frac{1}{2}(\partial_{+}g)^{2}-2e^{g}\bigg].
\ee
We may introduce new coordinates $x,y$ by defining
\ba
\la{3.22}
& x = u-\tau_{-}\cos u, \qquad y = \tau_{+}\cos u, \qquad \beta=\frac{2u}{\cos u}, \
\ea
where $x\in[-u,u]$ and $y\in[0,4u]$.
Notice that using (\ref{1.5}), we have $\beta = 2\beta_{\rm DSSYK}$ and $u = \frac{\pi}{2}-\theta$.
In the new coordinates, the action reads
\be
\la{3.23}
S = \frac{1}{8\l}\int_{-u}^{u}dx\int_{0}^{4u}dy\bigg[-\frac{1}{2}(\partial_{x}g)^{2}+\frac{1}{2}(\partial_{y}g)^{2}-\frac{2}{\cos^{2}u}e^{g}\bigg].
\ee
%with equations of motion.
%\be
%\partial_{x}^{2}g-\partial_{y}^{2}g-\frac{2}{\cos^{2}u}e^{g}=0.
%\ee
The static classical solution independent of $y$ and obeying $g=0$ at $x=\pm u$ is
\be
g_{\rm cl}(x) = \log\frac{\cos^{2}u}{\cos^{2}x},\ \ \text{with classical action}\ \ 
S_{\rm cl} = \frac{2u}{\l}(u-2\tan u).
\ee
Perturbation theory in $\l$ is set up by writing $g(x,y) = g_{\rm cl}(x)+\eee(x,y)\,\sql$
and replacing into the action (\ref{3.23}). This gives
\bea
\la{3.26}
S &= S_{\rm cl}+S_{2}+S_{\rm int}, \qquad S_{2} = \frac{1}{8}\int dxdy\bigg[-\frac{1}{2}(\partial_{x}\eee)^{2}+\frac{1}{2}(\partial_{y}\eee)^{2}-\frac{1}{\cos^{2}u}\eee^{2}\bigg], \\
S_{\rm int} &= -\frac{1}{4}\sum_{n=3}^{\infty}\frac{\l^{\frac{n}{2}-1}}{n!}I_{n}, \qquad I_{n}=\int dx dy \frac{1}{\cos^{2}x}\eee(x,y)^{n}.
\eea

\paragraph{Propagator and its regularization}

The propagator of the $g(x,y)$ field obeys
\ba
\la{3.27}
& \frac{1}{8}\bigg(\partial_{x}^{2}-\partial_{y}^{2}-\frac{2}{\cos^{2}x}\bigg)\langle \eee(x,y)\eee(x',y')\rangle = \delta(x-x')\hat\delta(y-y'),\\
\la{3.28}
& \wh\delta(y) = \sum_{m\in\mathbb Z}\delta(y+4u\, m) = \frac{1}{4u}\sum_{n\in\mathbb Z}e^{2\pi i n \frac{y}{4u}} = \frac{1}{4u}\sum_{n\in\mathbb Z}
e^{i\wt n y}, \qquad \wt n = \frac{\pi n}{2u}.
\ea
The factorized Ansatz
\be
\la{3.29}
\langle \eee(x,y)\eee(x',y')\rangle = \frac{2}{u}\sum_{n\in\mathbb Z}D_{n}(x,x') e^{i\wt n (y-y')},
\ee
implies
\be
\la{3.30}
\bigg(\partial_{x}^{2}+\wt n^{2}-\frac{2}{\cos^{2}x}\bigg)D_{n}(x,x')= \delta(x-x').
\ee
In \cite{Okuyama:2023bch}, it is proved that for $x>x'$
\be
\la{3.31}
\langle \eee(x,y)\eee(x',y')\rangle = \frac{1}{u\tan u}\bigg[-\frac{f(x)f(-x')}{1+u\tan u}+\sum_{|n|\ge 1}(-1)^{n}\frac{f_{n}(x)f_{n}(-x')}{\wt n^{2}(\wt n^{2}-1)} e^{i\wt n (y-y')}\bigg],
\ee
where for the non-zero modes
\ba
f_{n}(x) &= f^{(1)}_{n}(x) f^{(2)}_{n}(u)-f^{(1)}_{n}(u)f^{(2)}_{n}(x), \\
f^{(1)}_{n}(x) &= \cos(\wt n x)\tan x-\wt n \sin(\wt n x), \qquad
f^{(2)}_{n}(x) = \sin(\wt n x)\tan x+\wt n \cos(\wt n x),
\ea
and for zero mode
\be
\la{3.34}
f(x) = \tan x (1+u\tan u)-\tan u(1+x \tan x), \qquad \bigg(\partial_{x}^{2}-\frac{2}{\cos^{2}x}\bigg)f(x) = 0.
\ee
%
%
%
%
%
%\separator
%
% we take a function  $f_{n}(x)$ which is a solution of the homogeneous equation with $f_{n}(u)=0$. Then, the propagator vanishing at $x=\pm u$ is given by 
%\ba
%D_{n}(x,x') &= \frac{1}{W_{n}}[f_{n}(x)f_{n}(-x')\,\theta(x-x')+f_{n}(-x)f_{n}(x')\,\theta(x'-x)],  \\
%W_{n} &= \{f_{n}(x), f_{n}(-x)\} = \partial_{x}f_{n}(x)f_{n}(-x)-\partial_{x}f_{n}(-x)f_{n}(x) = \text{constant}.
%\ea
%For $n\neq 0$, two independent solutions of (\ref{3.30}) are
%\be
%f^{(1)}_{n} = \cos(\wt n x)\tan x-\wt n \sin(\wt n x), \qquad
%f^{(2)}_{n} = \sin(\wt n x)\tan x+\wt n \cos(\wt n x),
%\ee
%and we can take
%\ba
%\la{3.39}
%f_{n}(x) &= f^{(1)}_{n}(x) f^{(2)}_{n}(u)-f^{(1)}_{n}(u)f^{(2)}_{n}(x), \qquad
%W_{n} = 2(-1)^{n}\, \wt n^{2}(\wt n^{2}-1)\, \tan u.
%\ea
%In the case $n=0$, two independent solutions are 
%\be
%f^{(1)}_{0} = \tan x, \qquad
%f^{(2)}_{0} = x\tan x+1.
%\ee
%We take again the combination (\ref{3.39}), \ie 
%\ba
%\la{x3.43}
%f_{0}(x) & \equiv f(x) = \tan x (1+u\tan u)-\tan u(1+x \tan x),\qquad
%W_{0} = -2\tan u(1+u\tan u).
%\ea
%Notice also that clearly
%\be
%\la{3x.44}
%\bigg(\partial_{x}^{2}-\frac{2}{\cos^{2}x}\bigg)f(x) = 0.
%\ee
%The propagator in (\ref{3.29}) is thus \underline{for $x>x'$}
%\be
%\la{3.31}
%\langle \eee(x,y)\eee(x',y')\rangle = \frac{1}{u\tan u}\bigg[-\frac{f(x)f(-x')}{1+u\tan u}+\sum_{|n|\ge 1}(-1)^{n}\frac{f_{n}(x)f_{n}(-x')}{\wt n^{2}(\wt n^{2}-1)} e^{i\wt n (y-y')}\bigg].
%\ee
The infinite sum in (\ref{3.31}) is divergent because $f_{n}$ is quadratic in $n$. This is taken into account in 
\cite{Okuyama:2023bch} by means of the following summation formulas (valid for  $|\varphi|<2u$)
\bea
\la{3.35}
J_{1}(\varphi) &= \sum_{|n|\ge 1}(-1)^{n}\frac{\cos(\wt n \varphi)}{\wt n^{2}(\wt n^{2}-1)} = -\frac{\varphi^{2}}{2}+\frac{2u^{2}}{3}+1-\frac{2u}{\sin(2u)}\cos\varphi, \\
J_{2}(\varphi) &= \sum_{|n|\ge 1}(-1)^{n}\frac{\sin(\wt n \varphi)}{\wt n(\wt n^{2}-1)} = -J_{1}'(\varphi), \qquad % \varphi-\frac{2u}{\sin(2u)}\sin\varphi, \\
J_{3}(\varphi) = \sum_{|n|\ge 1}(-1)^{n}\frac{\cos(\wt n \varphi)}{\wt n^{2}-1} =  -J_{1}''(\varphi), \\ %1-\frac{2u}{\sin(2u)}\cos\varphi, \\
J_{4}(\varphi) &= \sum_{|n|\ge 1}(-1)^{n}\frac{\wt n\sin(\wt n \varphi)}{\wt n^{2}-1} = J_{1}'''(\varphi), \qquad % -\frac{2u}{\sin(2u)}\sin\varphi, \\
J_{5}(\varphi) = \sum_{|n|\ge 1}(-1)^{n}\frac{\wt n^{2}\cos(\wt n \varphi)}{\wt n^{2}-1} = J_{1}''''(\varphi). %-\frac{2u}{\sin(2u)}\cos\varphi.
\eea
While the sums $J_{1}, \dots, J_{4}$ are  convergent, the last sum $J_{5}$ is actually divergent. This implies that the proposed relation to $J_{1}''''$ in (\ref{3.35})
is formal  \footnote{We thank K. Okuyama and K. Suzuki for clarifications.} and  should be regarded as a specific regularization prescription. 
Using (\ref{3.35}), one finds
\be
\la{3.36}
x>x'\qquad \langle \eee(x,y)\eee(x',y')\rangle_{\rm reg} = \frac{f(x)f(-x')}{1+u\tan u}.
\ee
Notice that this is independent of $y, y'$. We do not discuss the coincident-point limit here; this will become
relevant in the following section.

\paragraph{Matter operator 2-point function}

The normalized 2-point function of the operator with dimension $\Delta$ in (\ref{2.16}) is computed in the $G\Sigma$ formalism by 
\be
G_{2}(\zeta, u) = \langle e^{\frac{\Delta}{\l}g(\zeta, y_{0})}\rangle = \frac{1}{Z}\int \mc Dg\ e^{-S}\ e^{\frac{\Delta}{\l}g(\zeta,y_{0})}.
\ee
where $y_{0}$ is an arbitrary reference point and $\zeta$ was given in (\ref{1.11}) that we recall here for convenience
%\footnote{
%We use from (\ref{3.22}) $\beta = \frac{2u}{\cos u}$ ,  $\tau_{12} = \tau_{1}-\tau_{2}$, $\tau_{1,2}=\frac{\beta}{2}\pm\frac{\tau}{4}$. 
%Recall also $u=\frac{\pi}{2}-\theta$.
%}
\be
\zeta = \frac{\pi}{2}-\theta-\frac{\tau}{2}\sin\theta.
\ee
The perturbative calculation of $G_{2}$ is apparently straightforward, at least expanding in small $\Delta$.  Indeed, using  (\ref{3.26}), we have 
%\be
%Z = e^{-S_{\rm cl}}\int D\eee(x,y) \ e^{-S_{2}} \ e^{-S_{\rm int}},
%\ee
%and
\ba
G_{2} &= e^{\frac{\Delta}{\l}g_{\rm cl}(\zeta)}\frac{1}{Z}\int Dg\, e^{-S_{\rm cl}-S_{2}}\, e^{\frac{\Delta}{\sql}\eee(\zeta, y_{0})-S_{\rm int}} \lp
= \bigg(\frac{1}{\cos^{2}\zeta}\bigg)^{\Delta/\l}\frac{\langle[1+\frac{\Delta}{\sql}\eee(\zeta, y_{0})
+\frac{\Delta^{2}}{\l}\eee(\zeta,y_{0})^{2}+O(\Delta^{3})]e^{-S_{\rm int}}\rangle_{S_{2}}}{\langle e^{-S_{\rm int}}\rangle_{S_{2}}}
\ea
In particular, the first perturbative terms in small $\l$ in the evaluation of (\ref{1.6}) are
\ba
\la{3.40}
-\partial_{\Delta}G_{2}|_{\Delta=0} &= \frac{2}{\l}\log\cos\zeta-\frac{1}{\sql}\frac{\langle \eee(\zeta,y_{0})e^{-S_{\rm int}}\rangle}{\langle e^{-S_{\rm int}}\rangle}\lp
= -\frac{1}{24} \langle \eee(\zeta,y_{0})I_{3}\rangle+\bigg(\frac{1}{27648}\langle \eee(\zeta,y_{0})I_{3}\rangle\langle I_{3}^{2}\rangle
-\frac{1}{82944}\langle \eee(\zeta,y_{0})I_{3}^{3}\rangle+\frac{1}{2304}\langle\eee(\zeta,y_{0})I_{3}\rangle\langle I_{4}\rangle\lp
-\frac{1}{2304}\langle\eee(\zeta,y_{0})I_{3}I_{4}\rangle-\frac{1}{480}\langle\eee(\zeta,y_{0})I_{5}\rangle\bigg)\, \l+O(\l^{2}).
\ea
The multiple expectation values should be evaluated using Wick's theorem and integrating over $x$ and $y$.
Let us look at the $O(\l)$ correction. We need 
\be
\la{3.41}
-\frac{1}{24}\langle\eee(\zeta,y_{0})\int dxdy\frac{1}{\cos^{2}x}\eee(x,y)^{3}\rangle = -\frac{1}{8}\int dxdy\frac{ 1}{\cos^{2}x}
\langle\eee(\zeta,y_{0})\eee(x,y)\rangle\ \langle \eee(x,y)^{2}\rangle.
\ee
If we now replace everywhere $\langle\bullet\rangle\to \langle\bullet\rangle_{\rm reg}$ and use   (\ref{3.36}) we get 
\ba
\la{3.42}
-\frac{1}{8}& \int dxdy\frac{ 1}{\cos^{2}x}
\langle\eee(\zeta,y_{0})\eee(x,y)\rangle\ \langle \eee(x,y)^{2}\rangle \lp
= -\frac{1}{8}\frac{4u}{(1+u\tan u)^{2}}\bigg[
\int_{-u}^{\zeta} \frac{dx}{\cos^{2}x}\frac{f(x)f(-x)}{\cos^{2}x}f(\zeta)f(-x)+
\int_{\zeta}^{u} \frac{dx}{\cos^{2}x}\frac{f(x)f(-x)}{\cos^{2}x}f(x)f(-\zeta)\bigg]\lp
= -\frac{u^{2}}{4}(\tan^{2}\zeta-\zeta\tan\zeta)+O(u^{3}).
\ea
The $u^{2}$ factor implies a vanishing result when $u\to 0$ which would prevent reproducing the correct 1-loop
correction in (\ref{1.10}) which is non-zero for $u=0$.  This issue is handled in \cite{Okuyama:2023bch} by using  (\ref{3.27}) for the regularized propagator
and dropping $\partial_{y}$, reducing the computation of (\ref{3.41}) to the solution of a non-homogeneous differential equation in $x$.
While this procedure gives the correct result, as we show below, it calls for a clarification since $y$-independence is not compatible with the $y$-dependent right-hand side in (\ref{3.27}).

The clean procedure is to define the composite operator expectation value  $\langle\eps(x,y)^{2}\rangle$ from the smooth coinciding limit of the regularized propagator,
while  for $\int dy \langle \eee(\zeta,y_{0})\eee(x,y)\rangle$ we use 
the exact representation (\ref{3.31}) that projects onto the 0-mode
\be
\la{3.43}
\zeta > x:\qquad \int_{0}^{4u} dy\langle \eee(\zeta,y_{0})\eee(x,y)\rangle = -\frac{4}{\tan u}\frac{f(\zeta)f(-x)}{1+u\tan u}.
\ee
The calculation is then identical to the one in (\ref{3.42}), but the $\zeta$-dependent propagator is rescaled by $-1/(u \tan u)$ and this cancels the unwanted $u^{2}$.
This explains why the procedure in  \cite{Okuyama:2023bch} works. In more detail, for generic $x,x'$, the full integrated propagator -- including Heaviside functions -- is 
\be
\int_{0}^{4u} dy'\langle \eee(x,y)\eee(x',y')\rangle = -\frac{4}{\tan u}\frac{f(x)f(-x')}{1+u\tan u}\theta(x-x')-\frac{4}{\tan u}\frac{f(x')f(-x)}{1+u\tan u}\theta(x'-x).
\ee
Thus, using (\ref{3.34}), one gets
\ba
\la{3.45}
\frac{1}{8} & \bigg(\partial_{x}^{2}-\frac{2}{\cos^{2}x}\bigg)\int_{0}^{4u} dy'\langle \eee(x,y)\eee(x',y')\rangle = 
-\frac{\{f(x), f(-x)\}}{2\tan u(1+u\tan u)}\delta(x-x') = \delta(x-x'),
\ea
and this is equivalent to using  (\ref{3.27}) with the $\partial_{y}^{2}$ term dropped. Indeed, one can check that the direct calculation (\ref{3.42})
with the projected propagator gives the right result at any finite $u$.

In principle, assuming the above procedure is still valid at all loop-orders, one should treat similarly  all terms in (\ref{3.40})
producing an increasing number of integrals as in (\ref{3.42}) or reducing each term to the solution of a non-homogeneous differential equation with 
homogeneous part as in (\ref{3.45}).

In the next sections, we  present an alternative procedure devised directly for the
infinite temperature limit (\ie $\beta=0$ or $u=0$), where the above regularization subtleties do not arise.

\section{Operator Liouville equation}
\la{sec:liouville}

In this section we present the method proposed to overcome the difficulties 
discussed in Section~\ref{sec:overview}. 
The key idea is to work directly with the operator equations of motion for 
$\hat\ell(\mu)$, $\alpha(\mu)$ and $\bar\alpha(\mu)$ in the DSSYK $q$-algebra, 
rather than with the integral expression (\ref{2.23}) or the bilocal 
correlator (\ref{1.6}). 
As we will show, these equations take a Liouville form, 
cf. (\ref{4.13}), and admit a systematic singular expansion in $\l$ 
that is algorithmically solvable to any loop order, as detailed 
in Section~\ref{sec:singular}.

The Krylov complexity (\ref{2.19}) in chord space can be written as the 1-point function of the Heisenberg Krylov number operator
\be
\la{4.1}
C(t) = \mmm{0}{\hat\ell(t)}{0}, \qquad \hat \ell(t) = e^{iHt}\hat\ell e^{-iHt}.
\ee
Taking into account the expression of the Hamiltonian in (\ref{2.4}), it is convenient to introduce
\be
\la{4.2}
\mu = \frac{iJt}{\sql},
\ee
define the operator
\be
\la{4.3}
U\equiv U(\mu) = e^{-\mu(\alpha+\bar\alpha)},
\ee
and consider
\be
\la{4.4}
\alpha(\mu) = U^{-1}\, \alpha\, U,
\ee
with a similar definition for $\bar\alpha$ and $\hat\ell$. The equation of motion of $\alp(\mu)$ is 
\be
\alpha'(\mu) = -U^{-1}\, [\alpha, \alpha+\bar\alpha]\, U.
\ee
Using the $q$-algebra relations we have 
\be
[\alpha, \bar\alpha] = \alpha\bar\alpha-\bar\alpha\alpha = 1+q\bar\alpha\alpha-\bar\alpha\alpha = 1-(1-q)\bar\alpha\alpha,
\ee
and thus
\be
\alpha'(\mu) = -1+(1-q)\, [\hat\ell(\mu)]_{q} = -1+(1-q)\frac{1-q^{\hat\ell(\mu)}}{1-q} = -q^{\hat\ell(\mu)}.
\ee
From the $\mu$-invariance of $\alpha+\bar\alpha$ (which commutes with $U$) we also have 
\be
\bar\alpha'(\mu) = -\alpha'(\mu) = q^{\hat\ell(\mu)}.
\ee
This is consistent with $\alpha+\bar\alpha$ being conserved (constant as $\mu$ is varied)
\be
\la{4.9}
\partial_{\mu}(\alpha(\mu)+\bar\alpha(\mu))=0.
\ee
Finally, for the operator $\hat\ell$ we obtain, using again (\ref{2.5}),
\be
\la{4.10}
\hat\ell'(\mu) = -U^{-1}\, [\hat\ell, \alpha+\bar\alpha]\, U  = \alpha(\mu)-\bar\alpha(\mu).
\ee
It follows from $\partial_{\mu}[\alpha(\mu)+\bar\alpha(\mu)]=0$ that
\be
\la{4.11}
\partial_{\mu}^{n}\hat\ell(\mu) = \partial_{\mu}^{n-1}[\alpha(\mu)-\bar\alpha(\mu)] = 2\partial_{\mu}^{n-1}\alpha(\mu), \qquad n\ge 2.
\ee
The equations of motion for the three operators $\alpha, \bar\alpha, \hat \ell$ 
\ba
\la{4.12}
\alpha'(\mu) &=  -q^{\hat\ell(\mu)}, \qquad \bar\alpha'(\mu) = q^{\hat\ell(\mu)}, \qquad
\hat\ell'(\mu) =  \alpha(\mu)-\bar\alpha(\mu),
\ea
imply the following (operator) one-dimensional Liouville equation and boundary conditions
\bea
\la{4.13}
& \hat\ell''(\mu) = -2q^{\hat\ell(\mu)}, \\
& \hat\ell(0) = \hat\ell, \qquad \hat\ell'(0) = \alpha-\bar\alpha.
\eea

\subsection{A toy analogy: the classical equation}

The equations (\ref{4.12}) and (\ref{4.13}) are non-trivial because they involve non-commuting operators. Nevertheless,  it is instructive to see what one gets 
treating $\alpha, \bar\alpha, \hat\ell$ as c-numbers and solving the ordinary differential equations
\ba
& \alp'(\mu) = -q^{\ell(\mu)}, \qquad \bar \alp'(\mu) = q^{\ell(\mu)}, \qquad \ell'(\mu) = \alp(\mu)-\bar \alp(\mu), \\
& \alp(0) =\alp, \qquad \bar \alp(0) = \bar \alp, \qquad \ell(0) = \ell.
\ea
The equation for $\ell(\mu)$ is again a one-dimensional Liouville equation
\be
\ell''(\mu) = -2q^{\ell(\mu)},
\ee
with general solution 
\be
\ell(\mu) = \frac{2}{\l}\log\bigg[\frac{1}{\sql c_{1}}\cos(\l (c_{1}\mu+c_{2}))\bigg]
\ee
The constants $c_{1},c_{2}$ are fixed by the boundary condition $\ell(0) = \ell$ and $\ell'(0) = \alp-\bar\alp \equiv D$.
Replacing $\mu = \tau/\sql$ and expanding in small $\l$, we get 
\be
\ell(\tau) = \frac{2}{\l}\log\cos\tau-D\, \tan\tau\,\frac{1}{\sql}+\ell(1+\tau\tan\tau)+\frac{1}{4}D^{2}(\tau\tan\tau-\tan^{2}\tau)+O(\sql).
\ee
If we assume that this structure survives in the solution of the operator equation and take the matrix element on the 0-chord state \footnote{We use $\mmm{0}{\hat\ell}{0}=0$, $\mmm{0}{D}{0}=0$
and $\mmm{0}{D^{2}}{0}\neq 0$.}
we find that  the classical contribution  in (\ref{1.13}) is reproduced, after analytic continuation $\tau=it$. The 1-loop term is also reproduced qualitatively in the 
sense that we have a combination of $\tau\tan\tau$ and $\tan^{2}\tau$.
This is encouraging and motivates a deeper analysis of the operator equation (\ref{4.13}).

\section{Series expansion of the operator Liouville equation at $\mu=0$}
\la{sec:taylor}

In this section, we examine the high-order Taylor expansion of (\ref{4.1}) at $\mu=0$. Using the variable $\mu$ in (\ref{4.2})
and exploiting (\ref{4.11}), we have 
\be
\la{5.1}
\hat\ell(\mu) = \hat\ell+(\alpha-\bar\alpha)\, \mu-q^{\hat\ell}\, \mu^{2}+2\sum_{n=3}^{\infty}\frac{\mu^{n}}{n!}\alpha^{(n-1)}(0),
\ee
where we use the boundary conditions in (\ref{4.13}). The second derivative $\alpha''(\mu)$ 
is easily computed by exploiting the operator relation (\ref{2.9})
\ba
\alpha''(\mu) = (-q^{\hat\ell(\mu)})' = \{(1-q)[\hat\ell(\mu)]_{q}-1\}' = (1-q)(\bar\alpha(\mu)\alpha(\mu))',
\ea
and, using (\ref{4.12}), we obtain
\be
\la{5.3}
\alpha''(\mu) = (1-q)[\bar\alpha(\mu)\alpha'(\mu)-\alpha'(\mu)\alpha(\mu)].
\ee
The next derivative is computed as
\ba
\alpha'''(\mu) &=  (1-q)(\bar\alpha\alpha'-\alpha'\alpha)' = (1-q)(\bar\alpha\alpha''-\alpha''\alpha-2(\alpha')^{2}).
\ea
Substituting $\alpha''$ from (\ref{5.3}), this gives
\ba
\alpha''' &= (1-q)[\bar\alpha (1-q)(\bar\alpha\alpha'-\alpha'\alpha)-(1-q)(\bar\alpha\alpha'-\alpha'\alpha)\alpha-2(\alpha')^{2}] \lp
= (1-q)[\bar\alpha (1-q)(-\bar\alpha q^{\hat\ell}+q^{\hat\ell}\alpha)-(1-q)(-\bar\alpha q^{\hat\ell}+q^{\hat\ell}\alpha)\alpha-2 q^{2\hat\ell}] \lp
=   -2(1-q)\,q^{2\hat\ell}-(1-q)^{2}\bar\alpha^{2}\, q^{\hat\ell}-(1-q)^{2}q^{\hat\ell}\alpha^{2}+2(1-q)^{2}\bar\alpha\, q^{\hat\ell}\, \alpha.
\ea
The procedure can be  continued algorithmically. Once we have the required number of terms in (\ref{5.1}), we take the matrix element
on the 0-chord state which is easily computed by exploiting the $q$-oscillator algebra.
%
%
%; it is convenient to 
%normal-order each term with all function of $\hat\ell$ to the left, then powers of $\bar\alpha$ then powers of $\alpha$,
%using systematically (\ref{2.10}) and 
%\be
%\alpha\bar\alpha = 1+q\bar\alpha\alpha = 1+q[\hat\ell]_{q} = [\hat\ell+1]_{q}.
%\ee

\subsection{Structure of Krylov complexity and variance}
\la{sec:series}

At this point we can evaluate the matrix element of the evolved operator in (\ref{5.1})
on the zero-chord state, \ie Krylov spread complexity. This gives \footnote{
The fact that the expansion is in even powers of $\mu$ is a general feature of spread complexity \cite{Muck:2026top}.
}
\ba
\la{5.6}
C &= \mmm{0}{\hat\ell(\mu)}{0} = -\mu ^2-\frac{1}{6} (1-q) \,\mu ^4-\frac{1}{180} (1-q)^2 (5+3 q)\, \mu^6\lp
-\frac{1}{10080}(1-q)^3 (35+56 q+35 q^2+10 q^3)\, \mu^8\lp
-\frac{1}{907200}(1-q)^4 (294+840 q+1134 q^2+951 q^3+525 q^4+189 q^5+35 q^6)\, \mu ^{10}+\cdots.
\ea
This expression is exact in $q$. Substituting $\mu = iJt/\sql$ and expanding in small $\l$ gives 
\be
\la{5.7}
C = K_{0}(J\,t)\, \frac{1}{\l}+K_{1}(J\,t)+K_{2}(J\,t)\, \l+K_{3}(J\,t)\, \l^{2}+\cdots,
\ee
where the functions $K_{n}$ have series expansions
\bea
\la{5-loops}
K_{0}(x) &= x^2-\frac{x^4}{6}+\frac{2 x^6}{45}-\frac{17 x^8}{1260}+\frac{62 
x^{10}}{14175}-\frac{691 x^{12}}{467775}+\cdots, \\
%%%%
K_{1}(x) &= \frac{x^4}{12}-\frac{11 x^6}{180}+\frac{x^8}{28}-\frac{268 
x^{10}}{14175}+\frac{8849 x^{12}}{935550}+\cdots, \\
%%%%
K_{2}(x) &= -\frac{x^4}{36}+\frac{11 x^6}{216}-\frac{547 x^8}{10080}+\frac{7723 
x^{10}}{170100}-\frac{295873 x^{12}}{8981280}+\cdots, \\
%%%%
K_{3}(x) &= \frac{x^4}{144}-\frac{23 x^6}{720}+\frac{35 x^8}{576}-\frac{214891 
x^{10}}{2721600}+\frac{1227467 x^{12}}{14968800}+\cdots, \\
%%%%
K_{4}(x) &= -\frac{x^4}{720}+\frac{529 x^6}{32400}-\frac{2777 
x^8}{50400}+\frac{999281 x^{10}}{9072000}-\frac{174980917 
x^{12}}{1077753600}+\cdots, \\
%%%%%
K_{5}(x) &= \frac{x^4}{4320}-\frac{17 x^6}{2400}+\frac{685 
x^8}{16128}-\frac{169843 x^{10}}{1306368}+\frac{971737043 
x^{12}}{3592512000}+\cdots.
\eea
and so on. An educated guess for the first two functions is 
\bea
\la{5.12}
K_{0}(x) &= 2\log\cosh x, \\
K_{1}(x) &= \frac{x}{4}\tanh x-\frac{1}{4}\tanh^{2}x,
\eea
but it seems hard to make a similar guess for the next terms. In the next section, we will solve this problem by
presenting a method to systematically derive  the closed-form expressions for all functions $K_{n}(x)$.

The explicit form of the operator $\hat\ell(\mu)$ in (\ref{5.1}) makes it easy to evaluate more complicated matrix elements. For instance, for the squared
Krylov number, we obtain 
\ba
& \mmm{0}{\hat\ell(\mu)^{2}}{0} = -\mu ^2+\frac{1}{3} (1+2 q) \mu ^4+\frac{1}{90} (1-q) (5+16 q+9 q^2) 
\mu ^6\lp
\qquad +\frac{1}{5040}(1-q)^2 (35+161 q+217 q^2+135 q^3+40 q^4) \mu^8 \lp
\qquad \frac{1}{453600}(1-q)^3 (294+1806 q+4032 q^2+5109 q^3+4254 q^4+2394 
q^5+896 q^6+175 q^7)\mu ^{10}+\cdots,
\ea
and thus the Krylov variance
\ba
\la{5.14}
& C^{(2)}\equiv \mmm{0}{\hat\ell^{2}}{0}-\mmm{0}{\hat\ell}{0}^{2} = -\mu ^2-\frac{2}{3} (1-q) \mu ^4-\frac{1}{90} (1-q)^2 (25+9 q) 
\mu ^6\lp
-\frac{1}{5040}(1-q)^3 (385+392 q+175 q^2+40 q^3)\,\mu ^8\lp
-\frac{1}{453600}(1-q)^4 (7056+12810 q+11928 q^2+7719 q^3+3465 
q^4+1071 q^5+175 q^6)\,\mu^{10}+\cdots.
\ea
In terms of the Lorentzian time, we have the small $\l$ expansion 
\be
\la{5.15}
C^{(2)} = K^{(2)}_{0}(J\,t)\frac{1}{\l}+K^{(2)}_{1}(J\, t)+K^{(2)}_{2}(J\,t)\,\l+\cdots,
\ee
where the functions $K^{(2)}_{n}(x)$ have Taylor expansion 
\bea
\la{var-exp}
K^{(2)}_{0}(x) &= x^2-\frac{2 x^4}{3}+\frac{17 x^6}{45}-\frac{62 x^8}{315}+\frac{1382 x^{10}}{14175}+\cdots, \\
%%%%
K^{(2)}_{1}(x) &= \frac{x^4}{3}-\frac{43 x^6}{90}+\frac{235 x^8}{504}-\frac{21067 x^{10}}{56700}+\cdots, \\
%%%%
K^{(2)}_{2}(x) &= -\frac{x^4}{9}+\frac{10 x^6}{27}-\frac{3259 x^8}{5040}+\frac{547873 
x^{10}}{680400}+\cdots , \\
%%%%
K^{(2)}_{3}(x) &= \frac{x^4}{36}-\frac{79 x^6}{360}+\frac{581 x^8}{864}-\frac{1752257 
x^{10}}{1360800}+\cdots.
\eea
A natural guess for the first expansion is 
\be
K^{(2)}_{0}(x) = \tanh^{2}x,
\ee
but, again, it is hard to make a similar Ansatz for the higher functions. This will be done in the next sections.

\subsection{Generating function for higher cumulants}

Higher order cumulants are defined as usual as 
\be
\la{5.20}
C^{(p)} = \frac{d^{p}}{dz^{p}}\log\mmm{0}{e^{z \hat\ell(\mu)}}{0}|_{z=0}.
\ee
In particular, with the notation $\ell_{n} \equiv \mmm{0}{\hat\ell^{n}}{0}$, 
the cases $p=2$ (variance) and $p=3$ read \footnote{For $p=2,3$, the cumulant $C^{(p)}$ coincides with the central moment $\mmm{0}{(\hat\ell-\ell_{1})^{p}}{0}$, 
but not for $p\ge 4$.
}
\be
C^{(2)} = \ell_{2}-\ell_{1}^{2}, \qquad C^{(3)} = \ell_{3}-3\ell_{1}\ell_{2}+2\ell_{1}^{3}.
\ee
If we are interested in expansions at some fixed order in $\mu$, as in (\ref{5.6}) and (\ref{5.14}), and want to explore the dependence on the
cumulant order, it is convenient to avoid using the expansion of $\hat\ell(\mu)$ and use a different alternative method.
We start from
\be
f(z; \mu) \equiv \mmm{0}{e^{z\hat\ell(\mu)}}{0} = \mmm{0}{U^{-1}e^{z\hat\ell}U}{0},\qquad U=e^{-\mu(\alpha+\bar\alpha)}.
\ee
Inserting a complete set of chord states, we have (for real $\mu$)
\be
\la{5.23}
f(z; \mu) = \sum_{\ell=0}^{\infty}e^{z\ell}\mmm{0}{U^{-1}}{\ell}\,\mmm{\ell}{U}{0} = \sum_{\ell=0}^{\infty}e^{z\ell}g_{\ell}(\mu)g_{\ell}(-\mu),
\ee
where
\be
\la{5.24}
g_{\ell}(\mu) = \mmm{\ell}{U^{-1}}{0} = \mmm{\ell}{\sum_{n=0}^{\infty}\frac{\mu^{n}}{n!}(\alpha+\bar\alpha)^{n}}{0}.
\ee
Expanding (\ref{5.24}) at any desired order $N$ in $\mu$, and using (\ref{2.7}), we see that the infinite sum in (\ref{5.23}) is  zero for $\ell>N$
and reduces to a finite sum. Explicitly, at order $\mu^{6}$,  we get for the logarithm of $f(z; \mu)$ 
\ba
\log f(z; \mu) &=(1-e^z) \mu ^2+\frac{1}{12} (1-4 e^z+3 e^{2 z}) (q-1) \mu^4\lp
-\frac{1}{360} (e^z-1) (q-1)^2 [5+q+10 e^{2 z} (4+q)-5 e^z (7+q)]
\mu ^6+\cdots.
\ea
Using now (\ref{5.20}) we get 
\ba
\la{5.26}
C^{(p)} &=- \mu ^2+\frac{1}{12} (-4 +3\cdot 2^{p}) (q-1) \mu^4\lp
-\frac{1}{360} (q-1)^2 [40-75\cdot 2^p+40\cdot 3^p+(6-15\cdot 2^p+10\cdot 3^p) q]\,\mu ^6+\cdots.
\ea
One can check that, taking $p=1,2$, this general expression  agrees with (\ref{5.6}) and (\ref{5.14}).

\section{Singular expansion of the operator Liouville equation}
\la{sec:singular}

Let us redefine $\mu\to \mu/\sql$ in the Liouville equation (\ref{4.13}). This gives
\ba
\la{6.1}
& \hat\ell''(\mu) = -\frac{2}{\l}e^{-\l\,\hat\ell(\mu)}, \\
\la{6.2}
& \hat\ell(0) = \hat\ell, \qquad \hat\ell'(0) = \frac{1}{\sql}(\alpha-\bar\alpha).
\ea
Since we want to build an expansion around $\l=0$, it is necessary to expand the $\alp, \bar\alp$ operators around their $\l=0$ limit which are
standard oscillators $a, \bar a$. From 
\be
\alpha\ket{\ell} = [\ell]_{q}^{1/2}\ket{\ell-1} = \sqrt\frac{[\ell]_{q}}{\ell}\sqrt{\ell}\ket{\ell-1},
\ee
we obtain 
\ba
\la{6.4}
\alpha &= a\, \sqrt\frac{[\hat\ell]_{q}}{\hat\ell} = \sqrt\frac{[\hat\ell+1]_{q}}{\hat\ell+1}\, a, \qquad
\bar\alpha = \sqrt\frac{[\hat\ell]_{q}}{\hat\ell}\,\bar a  = \bar a \,\sqrt\frac{[\hat\ell+1]_{q}}{\hat\ell+1}, \qquad
\hat\ell = \bar a a,
\ea
and we have again the shift identities
\be
a\, f(\hat \ell) = f(\hat\ell+1)\, a, \qquad
\bar a \, f(\hat \ell) = f(\hat\ell-1)\, \bar a .
\ee
Using (\ref{6.4}) the second boundary condition in (\ref{6.2}) is  
\be
\la{6.6}
\hat\ell'(0) = \frac{1}{\sql}\bigg(a\, \sqrt\frac{[\hat\ell]_{q}}{\hat\ell}-\sqrt\frac{[\hat\ell]_{q}}{\hat\ell}\, \bar a \bigg).
\ee
Guided by (\ref{5.7}), we now write the small $\l$ expansion of $\hat\ell(\mu)$ in the form 
\be
\la{6.7}
\hat\ell(\mu) = X_{0}(\mu)\frac{1}{\l}+X_{1}(\mu)\frac{1}{\sql}+X_{2}(\mu)+X_{3}(\mu)\sql+X_{4}(\mu)\l+\cdots,
\ee
where $X_{n}(\mu)$ are operators that, in general, need not mutually commute. As we will see below, it is necessary to include in (\ref{6.7})
half-integer powers of $\l$. They will cancel in the expectation value on the 0-chord state. We also expand
the boundary condition for $\hat\ell'(0)$ in (\ref{6.6})
\ba
\la{6.8}
\hat\ell'(0) &= \frac{1}{\sql}\bigg(a\, \sqrt\frac{[\hat\ell]_{q}}{\hat\ell}-\sqrt\frac{[\hat\ell]_{q}}{\hat\ell}\, \bar a \bigg) 
= \frac{1}{\sql}(a-\bar a)+\frac{1}{4}\bigg(a\, (1-\hat \ell) - (1-\hat \ell)\,\bar a\bigg)\, \sql \lp
+\frac{1}{96}\bigg[
a\,(\hat\ell-1)\,(5\hat\ell-1)-(\hat\ell-1)\,(5\hat\ell-1)\,\bar a
\bigg]\,\l^{3/2}+\cdots
\ea

\subsection{Leading and next-to-leading contributions}

The leading-order equation obtained by expanding (\ref{6.1}) is  
\be
X_{0}''(\mu) = -2e^{-X_{0}(\mu)}, \qquad X_{0}(0) = X_{0}'(0) = 0,
\ee
with solution 
\be
X_{0}(\mu) = \log\cos^{2}\mu.
\ee
Since $X_{0}$ is a multiple of the identity, we get at the next order
\be
X_{1}''(\mu) = \frac{2}{\cos^{2}\mu}X_{1}(\mu),
\ee
with boundary condition
\be
X_{1}(0) = 0, \qquad X_{1}'(0) = a-\bar a.
\ee
The operator appears only in the second condition, so we can treat it as a c-number and get
\be
\la{6.13}
X_{1}(\mu) = (a-\bar a)\, \tan\mu.
\ee

\subsection{Higher order corrections}

\paragraph{Determination of $X_{2}(\mu)$}

We can expand the Liouville equation at the next order, taking into account that $X_{2}$ does not necessarily commute with $X_{1}$. 
To organize the calculation, it is convenient to  rewrite  (\ref{6.7}) as 
\be
\hat \ell(\mu) = \frac{1}{\l}\log\cos^{2}\mu+\frac{1}{\sql}\Omega, \qquad \Omega = X_{1}+ X_{2}\,\sql+ X_{3}\,\l+X_{4}\,\l^{3/2}+\cdots.
\ee
Since $\Omega$ is the only operator in the equation (\ref{6.1}), we can expand and get 
\ba
\la{6.15}
\Omega'' &= \frac{2}{\cos^{2}\mu}\Omega-\frac{1}{\cos^{2}\mu}\Omega^{2}\sql 
+\frac{1}{3\cos^{2}\mu}\Omega^{3}\,\l
-\frac{1}{12\cos^{2}\mu}\Omega^{4}\,\l^{3/2}+\cdots.
\ea
The $\l^{0}$ term is satisfied by (\ref{6.13}). The term $\sql$ gives 
 \be
X_{2}'' =  \frac{2}{\cos^{2}\mu}X_{2}-\frac{1}{\cos^{2}\mu}X_{1}^{2}.
\ee
The boundary condition from (\ref{6.2}) is  
\be
X_{2}(0) = \hat\ell, \qquad X_{2}'(0) = 0.
\ee
So we need to solve 
\bea
\la{6.18}
& X_{2}'' =  \frac{2}{\cos^{2}\mu}X_{2}-\frac{1}{\cos^{2}\mu}(a-\bar a)^{2}\tan^{2}\mu, \\
& X_{2}(0) = \hat\ell, \qquad X_{2}'(0) = 0.
\eea
At this point, it is useful to first consider the general equation 
\be
\la{6.19}
Y'' = \frac{2}{\cos^{2}\mu}Y+Z,\qquad Y(0) = A, \qquad Y'(0)=B,
\ee
where $Z,A,B$ are generic operators (not necessarily commuting). The homogeneous solution for $Z=0$ has the two independent solutions
\be
Y_{1}(\mu) = \tan\mu,\qquad Y_{2}(\mu) = \mu \tan\mu+1,
\ee
with Wronskian
\be
W = Y_{1}Y_{2}'-Y_{1}'Y_{2} = -1.
\ee
The solution of (\ref{6.19}) is then 
\ba
Y(\mu) &= \frac{(AY_{2}'(0)-BY_{2}(0))Y_{1}(\mu)+(BY_{1}(0)-A Y_{1}'(0))Y_{2}(\mu)}{W(0)}\lp
+\int_{0}^{\mu}d\mu'\ \frac{Y_{1}(\mu')Y_{2}(\mu)-Y_{1}(\mu)Y_{2}(\mu')}{W(\mu')}
Z(\mu'),
\ea
and in our case it reads
\bea
\la{6.23}
Y(\mu) &= A (\mu\tan\mu+1)+B\tan\mu+\int_{0}^{\mu}d\mu'\ G(\mu,\mu')\,Z(\mu'), \\
G(\mu,\mu') &= \tan\mu(\mu'\tan\mu'+1)-\tan(\mu')(\mu\tan\mu+1).
\eea
A key remark is that the three operators $Z, A, B$ do not mix in the sense that $Y(\mu)$ is linear in all of them, so their non-commutativity
is irrelevant.

Returning to problem (\ref{6.18}), we have 
\ba
A = \hat\ell, \qquad B=0, \qquad Z = -\frac{\tan^{2}\mu}{\cos^{2}\mu}(a-\bar a)^{2}.
\ea
Using
\ba
\int_{0}^{\mu}d\mu'\ G(\mu,\mu')\,\frac{\tan^{2}\mu'}{\cos^{2}\mu'} = \frac{1}{4}\tan\mu(-\mu+\tan\mu), 
\ea
we get the solution
\be
X_{2}(\mu) = \hat\ell(1+\mu\tan\mu)+\frac{1}{4}(a-\bar a)^{2}(\mu-\tan\mu)\tan\mu.
\ee

\paragraph{Determination of $X_{3}(\mu)$}

The next order is the  $O(\l)$ term in (\ref{6.15}). Reading boundary conditions from (\ref{6.8}) we need to solve
\ba
& X''_{3} = \frac{2}{\cos^{2}\mu}X_{3}+\frac{1}{\cos^{2}\mu}[-X_{1}X_{2}-X_{2}X_{1}]+\frac{1}{3\cos^{2}\mu}X_{1}^{3},\\
& X_{3}(0) = 0, \qquad X_{3}'(0) = \frac{1}{4}(a\, (1-\hat \ell)-(1-\hat\ell) \,\bar a).
\ea
Denoting $D=a-\bar a$, we now apply (\ref{6.23}) with
\ba
A &= 0, \qquad
B = \frac{1}{4}(a\, (1-\hat\ell)-(1-\hat\ell)\, \bar a), \\
Z &= -\frac{\tan\mu(1+\mu\tan\mu)}{\cos^{2}\mu}(D\,\hat\ell+\hat\ell\,D)+\frac{(5\tan\mu-3\mu)\tan^{2}\mu}{6\cos^{2}\mu}D^{3}.
\ea
We need the new functions
\ba
\int_{0}^{\mu} & d\mu'\ G(\mu,\mu')\,\frac{\tan\mu'(1+\mu'\tan\mu')}{\cos^{2}\mu'} = \frac{1}{4}\bigg(\frac{\mu}{\cos^{2}\mu}-\tan\mu\bigg).
\ea
\ba
\int_{0}^{\mu} & d\mu'\ G(\mu,\mu')\,\frac{(5\tan\mu'-3\mu')\tan^{2}\mu'}{6\cos^{2}\mu'} = 
-\frac{1}{8}\frac{\mu}{\cos^{2}\mu}+\frac{1}{32}\tan\mu
+\frac{3}{32}\frac{\tan\mu}{\cos^{2}\mu}-\frac{1}{96}\tan^{3}\mu.
\ea
Thus
\ba
X_{3}(\mu) &= \frac{1}{4}(a\, (1-\hat\ell)-(1-\hat\ell)\,\bar a)\tan\mu-\frac{1}{4}\bigg(\frac{\mu}{\cos^{2}\mu}-\tan\mu\bigg)\,(D\,\hat\ell+\hat\ell D)\lp
+\bigg(-\frac{1}{8}\frac{\mu}{\cos^{2}\mu}+\frac{1}{32}\tan\mu
+\frac{3}{32}\frac{\tan\mu}{\cos^{2}\mu}-\frac{1}{96}\tan^{3}\mu\bigg)\,D^{3}.
\ea

\paragraph{Determination of $X_{4}(\mu)$}

Finally,  we need the term $O(\l^{3/2})$ in (\ref{6.15}). This gives a rather complicated expression for $Z$, but integration with $G(\mu, \mu')$ can be done in terms of 
elementary functions as in previous cases. We have $A=B=0$ and thus
\ba
X_{4}(\mu) = \int_{0}^{\mu} & d\mu'\ G(\mu,\mu') Z(\mu') = -\frac{1}{4}\mu(\mu\sec^{2}\mu+\tan\mu)\, \hat\ell^{2}\lp
+\frac{1}{16}\tan\mu(\mu-\tan\mu)(2D^{2}-a\hat\ell\,D-D\,a\,\hat\ell+D\,\hat\ell\bar a+\hat \ell\,\bar a\, D) \lp
+\frac{1}{48}(2+\mu\tan\mu-\sec^{2}\mu(2+3\mu^{2}-4\mu\tan\mu))\, (D^{2}\, \hat\ell+\hat\ell\, D^{2}) \lp
+\frac{1}{12}\tan\mu(\mu(1+\sec^{2}\mu)-2\tan\mu)\,D\hat\ell D\lp
+\frac{1}{64}(2-2\sec^{4}\mu+\mu\sec^{2}\mu(-\mu+4\tan\mu)+\mu\tan\mu)\,D^{4}.
\ea

\paragraph{Determination of $X_{5}(\mu), X_{6}(\mu), \dots$}

The procedure needed to go up to 5-loop order is the same, but we need $X_{5}, \dots, X_{10}$. The calculation is  completely algorithmic and is 
easily implemented in a computer algebra system. 
%As an illustration,  the expression for $X_{5}$ is 
%\ba
%X_{5} &= f_{1}(\mu)\, D^{4}+f_{2}(\mu)\, \hat\ell\,D\,\hat\ell +f_{3}(\mu)\, D\, \hat\ell^{2}\lp
%+f_{4}(\mu)\,(
%4D^{3}-a\, \hat \ell\, D^{2}-D\, a\, \hat\ell\, D-D^{2}\, a\, \hat\ell+D^{2}\, \hat\ell\, \bar a+D\, \hat\ell\, \bar a\, D+\hat\ell\, \bar a\, D^{2})
%+f_{5}(\mu)\, (D^{3}\, \hat\ell+\hat\ell\, D^{3})\lp
%+f_{6}(\mu)\,(D^{2}\, \hat\ell\, D+D\, \hat\ell\, D^{2})+f_{7}(\mu)\, (2D\, \hat\ell+2\hat\ell\,D-a\, \hat\ell^{2}-\hat\ell\, a\, \hat\ell+\hat\ell\, \bar a\, \hat\ell+\hat\ell^{2}\, \bar a)\lp
%+f_{8}(\mu)\, (D\, \hat\ell^{2}-\hat\ell^{2}\, D)+\frac{1}{96}\tan\mu\, [a\, \hat\ell\, (-12+5\hat\ell)-\hat\ell\, (-12+5\hat\ell)\, \bar a],
%\ea
%with
%\bea
%f_{1}(\mu) &= \frac{1}{640}[-5 \mu (1+\tan^{2}\mu)(3-2\mu \tan \mu+4\tan^{2}\mu)+\tan \mu(15+20\tan^{2}\mu+8\tan^{4}\mu)], \\
%f_{2}(\mu) &=\frac{1}{12}[2\tan \mu-\mu(2+\tan^{2}\mu)+\mu^{2}\tan \mu(1+\tan^{2}\mu)] , \\
%f_{3}(\mu) &= \frac{1}{24}[-\tan \mu+\mu(1-\tan^{2}\mu+4\mu\tan \mu(1+\tan^{2}\mu))], \\
%f_{4}(\mu) &= \frac{1}{96}[-3\mu(1+\tan^{2}\mu)+\tan \mu(3+2\tan^{2}\mu)], \\
%f_{5}(\mu) &= \frac{1}{192}[\tan \mu(9+2\tan^{2}\mu)+\mu(-9-13\tan^{2}\mu-6\tan^{4}\mu+8 \mu\tan \mu (1+\tan^{2}\mu))], \\
%f_{6}(\mu) &= \frac{1}{192}[\tan \mu(9+2\tan^{2}\mu)+\mu(-9-17\tan^{2}\mu-6\tan^{4}\mu+4\mu\tan \mu(1+\tan^{2}\mu))], \\
%f_{7}(\mu) &= \frac{1}{16}[-\tan \mu+\mu(1+\tan^{2}\mu)], \\
%f_{8}(\mu) &=-\frac{1}{48}[-\tan \mu+\mu(1-\tan^{2}\mu+4\mu\tan \mu(1+\tan^{2}\mu))] .
%\eea
%The expression for $X_{6}$ has a similar structure.

\section{Results}
\la{sec:results}

\subsection{Krylov complexity at 5-loops}

The Krylov complexity is given by the matrix element  
\be
\mmm{0}{\hat\ell(\mu)}{0} = \mmm{0}{\frac{1}{\l}X_{0}(\mu)+\frac{1}{\sql}X_{1}(\mu)+X_{2}(\mu)+\cdots}{0},
\ee
where we keep terms up to $X_{10}(\mu)$. Their expressions are polynomial in $\hat\ell$ and $a, \bar a$,
reducing the matrix element evaluation to standard oscillator algebra manipulations. In particular, the expectation value of all odd functions $X_{2n+1}$ vanishes, 
so that the series runs in integer powers of $\l$. As an example for the case of even functions, we have 
\ba
\mmm{0}{X_{2}(\mu)}{0} &= \mmm{0}{[\hat\ell(1+\mu\tan\mu)+\frac{1}{4}(a-\bar a)^{2}(\mu-\tan\mu)\tan\mu]}{0}\lp
= \frac{1}{4}\tan^{2}\mu-\frac{1}{4}\mu\tan\mu,
\ea
where we used $\mmm{0}{\hat\ell}{0}=0$ and $\mmm{0}{(a-\bar a)^{2}}{0} = -\mmm{0}{a\bar a+\bar a a}{0} = -\braket{1}{1} =  -1$.
Working  out the next terms up to $X_{10}$,  we get 
\ba
& \mmm{0}{\hat\ell(\mu)}{0} = 2\log\cos\mu\,\frac{1}{\l}+\frac{1}{4}\tan^{2}\mu-\frac{1}{4}\mu\tan\mu \lp
+\bigg[-\frac{3}{64}\mu^{2}+\frac{13}{192}\mu \tan\mu-\bigg(\frac{1}{48}+\frac{3}{64}\mu^{2}\bigg)\tan^{2}\mu
+\frac{5}{48}\mu\tan^{3}\mu-\frac{3}{32}\tan^{4}\mu
\bigg]\, \l \lp
+\bigg[
\frac{31}{1536}\mu^{2}+\bigg(-\frac{55}{1536}+\frac{1}{768}\mu^{2}\bigg)\, \mu\, \tan \mu+\bigg(\frac{1}{64}+\frac{109}{1536}\mu^{2}\bigg)\, \tan^{2}\mu \lp
+\bigg(-\frac{109}{768}+\frac{1}{768}\mu^{2})\, \mu\, \tan^{3}\mu+\bigg(\frac{5}{64}+\frac{13}{256}\mu^{2}\bigg)\, \tan^{4}\mu
-\frac{7}{64}\mu\, \tan^{5}\mu+\frac{5}{64}\tan^{6} \mu
\bigg]\, \l^{2}\lp
+\bigg[-\frac{685 \mu ^2}{147456}-\frac{41 \mu^4}{24576}+\bigg(\frac{7201}{737280}+\frac{11 \mu^2}{12288}\bigg) \mu \tan \mu 
+\bigg(-\frac{59}{11520}-\frac{11845 \mu ^2}{147456}-\frac{41 \mu^4}{6144}\bigg) \tan^2 \mu \lp
+\bigg(\frac{3001}{18432}+\frac{131 \mu^2}{12288}\bigg)\,\mu \tan ^3 \mu
+\bigg(-\frac{637}{7680}-\frac{1055 \mu^2}{6144}-\frac{41 \mu ^4}{8192}\bigg) \tan^4\mu
+\bigg(\frac{5051}{15360}+\frac{5 \mu^2}{512}\bigg)\mu \tan^5\mu \lp
+\bigg(-\frac{45}{256}-\frac{295 \mu ^2}{3072}\bigg) \tan^6\mu
+\frac{45}{256} \mu  \tan ^7 \mu-\frac{105}{1024} \tan^8\mu\bigg]\,\l^{3}+O(\l^{4}),
\ea
where we do not write the 5-loop contribution for brevity. It will be explicitly given in  (\ref{1.15}) below.
After substitution $\mu = i J t$,  the first line confirms (\ref{5.12}). The next lines give contributions up to 5-loops according to 
\ba
K_{0}(x) &= 2\log\cosh x, \qquad K_{1}(x) = \frac{1}{4}x\tanh x-\frac{1}{4}\tanh^{2}x, \notag \\
\la{1.15}
K_{2}(x) &= \frac{3}{64}x^{2}-\frac{13}{192}x \tanh x+\frac{1}{192}(4-9x^{2})\tanh^{2}x+\frac{5}{48}x \tanh^{3}x-\frac{3}{32}\tanh^{4}x, \\
K_{3}(x) &= -\frac{31}{1536}x^{2}+\frac{1}{1536}(55+2x^{2})x \tanh x+\frac{1}{1536}(-24+109 x^{2})\tanh^{2}x \lp
-\frac{1}{768}(109+x^{2})x \tanh^{3}x+\frac{1}{256}(20-13x^{2})\tanh^{4}x
+\frac{7}{64}x \tanh^{5}x-\frac{5}{64}\tanh^{6}x.\notag \\
%%%%%
K_{4}(x) &= -\frac{1}{147456}x^2 (-685+246 x^2)+\frac{1}{737280} (-7201+660 x^2)x \tanh (x)\lp
+\frac{1}{737280}(3776-59225 x^2+4920 x^4) \tanh^2 x
-\frac{1}{36864} (-6002+393 x^2) x\tanh^3 x\lp
+\frac{1}{122880}(-10192+21100 x^2-615 x^4) \tanh^4 x 
+\frac{1}{15360} (-5051+150 x^2) x\tanh^5 x \lp
-\frac{5}{3072} (-108+59 x^2) \tanh^6 x 
+\frac{45}{256} x \tanh ^7 x-\frac{105}{1024} \tanh ^8 x, \notag \\
%%%%%
K_{5}(x) &= \frac{11}{5898240} x^2 (-3271+1700 x^2)
-\frac{1}{5898240} (-72461+85230 x^2+14544 x^4) x\tanh x \lp
+\frac{1}{5898240}(-36480+716051 x^2-116200 x^4) \tanh^2 x \lp
+\frac{1 }{2949120} (-614081+220815 x^2+18180 x^4) x\tanh^3 x\lp
+\frac{1}{196608}(19328-89237 x^2+5550 x^4) \tanh^4 x
-\frac{1}{81920}(-61270+9050 x^2+303 x^4) x\tanh^5 x \lp
+\frac{1}{49152}(-17408+28822 x^2-575 x^4) \tanh^6 x
+\frac{1}{4096} (-3801+205 x^2) x\tanh^7 x \lp
-\frac{35}{4096} (-52+29 x^2) \tanh^8 x
+\frac{385}{1024} x \tanh^9 x
-\frac{189}{1024} \tanh^{10} x.
\ea
One can check that they reproduce the non-trivial expansions in (\ref{5-loops}).

\subsection{Krylov variance and third cumulant}

From the expansion that we computed, we can determine $C^{(2)}$ at 4-loop order
 and $C^{(3)}$ at 3-loop  order. 
 %For brevity, we present the results at 3-loop and 2-loop respectively. \footnote{The full expressions are available from the authors on request.}
A straightforward calculation gives (we set $J=1$ for simplicity)
\ba
\ell_{2} &= 4 \log^2\cos\mu\,\frac{1}{\l^{2}}+\bigg[-\tan^{2}\mu+\log\cos\mu\,\tan\mu\,(-\mu+\tan\mu)\bigg]\frac{1}{\l}\lp
-\frac{1}{4}\mu\tan\mu+\bigg(\frac{1}{4}+\frac{3\mu^{2}}{16}\bigg)\, \tan^{2}\mu-\frac{5}{8}\mu\tan^{3}\mu+\frac{11}{16}\tan^{4}\mu\lp
+\log\cos\mu\bigg[-\frac{3\mu^{2}}{16}+\frac{13}{48}\mu\tan\mu+\bigg(-\frac{1}{12}-\frac{3\mu^{2}}{16}\bigg)\tan^{2}\mu+\frac{5}{12}\mu\tan^{3}\mu-\frac{3}{8}\tan^{4}\mu+O(\l),
\ea
where for brevity we wrote only the terms up to $O(\l^{0})$. The  Krylov variance takes the form (\ref{5.15}) with the following exact  functions
\ba
K_{0}^{(2)}(x) &= \tanh^{2}x, \notag \\
\la{1.18}
K_{1}^{(2)}(x) &= \frac{1}{4}x \tanh x+\frac{1}{8}(-2+x^{2})\tanh^{2}x-\frac{1}{2}x \tanh^{3}x+\frac{5}{8}\tanh^{4}x, \\
K_{2}^{(2)}(x) &= \frac{7}{64}x^{2}-\frac{1}{192}x(37+6x^{2})\tanh x+\frac{1}{48}(4-21x^{2})\tanh^{2}x+\frac{1}{192}x(199+6x^{2})\tanh^{3}x \lp
+\frac{1}{64}(-44+19x^{2})\tanh^{4}x-\frac{13}{16}x \tanh^{5}x+\frac{2}{3}\tanh^{6}x. \notag  \\
%%%%%
K_{3}^{(2)}(x) &= \frac{1}{1536}x^2 (-103+12 x^2)
+\frac{1}{1536} (199+43 x^2)x \tanh x
+\frac{1}{256} (-16+171 x^2-11 x^4) \tanh ^2 x\lp
+\frac{1}{1536} (-2212+29 x^2) x\tanh ^3 x
+\frac{1}{256} (196-355 x^2+9 x^4) \tanh ^4 x\lp
+\frac{1}{768} (2267 -36 x^2) x\tanh ^5 x
+\frac{1}{256} (-440+199 x^2) \tanh^6 x
-\frac{103}{64} x \tanh ^7 x
+\frac{65}{64} \tanh ^8 x. \notag \\
%%%%%%
K_{4}^{(2)}(x) &= -\frac{1}{147456}x^2 (-6061+3780 x^2)
+\frac{1}{737280} (-60961+61500 x^2+14160 x^4)x \tanh x \lp
+\frac{1}{184320}(7664-200090 x^2+31635 x^4) \tanh^2 x\lp
-\frac{1}{737280} (-1475701+408180 x^2+38640 x^4) x\tanh^3 x \lp
+\frac{1}{49152}(-48128+208473 x^2-11844 x^4) \tanh^4 x
+\frac{1}{2048} (-15281+1817 x^2+68 x^4) x\tanh^5 x \lp
+\frac{1}{36864}(135872-206130 x^2+3501 x^4) \tanh^6 x
-\frac{1}{3072} (-29084+1281 x^2)x\tanh^7 x \lp
+\frac{1}{768} (-3668+1839 x^2) \tanh^8 x
-\frac{751}{192} x \tanh ^9 x+\frac{2589}{1280} \tanh ^{10} x.
\ea
Again, one can check that they reproduce the expansions in (\ref{var-exp}).
Finally, for the third-order cumulant, we similarly find the expressions 
\ba
\la{1.19}
K_{0}^{(3)}(x) &= \frac{1}{2}\tanh^{2}x\,(2+x\tanh x-3\tanh^{2}x), \\
K_{1}^{(3)}(x) &= \frac{1}{16}\tanh x\,[4x+(-4+15x^{2})\tanh x-x(51+2x^{2})\tanh^{2}x-3(-16+3x^{2})\tanh^{3}x\lp
+42 x\tanh^{4}x-44\tanh^{5}x].\notag \\
K_{2}^{(3)}(x) &= \frac{19 x^2}{64}-\frac{1}{384}  (218+81 x^2) x\tanh x
+\frac{1}{768} (208-2073 x^2+90 x^4) \tanh^2 x \lp
+\frac{1}{768}  (4967+216 x^2) x\tanh ^3 x+\frac{1}{256} (-984+1357 x^2-30 x^4) \tanh^4 x \lp
-\frac{1753}{128} x \tanh ^5 x+\frac{1}{128} (1176-383 x^2) 
\tanh ^6 x+\frac{123}{16} x \tanh ^7 x-\frac{351}{64} \tanh ^8 x. \notag, \\
%%%%%%
K_{3}^{(3)}(x) &= \frac{11 x^2 (-58+21 x^2)}{3072}
-\frac{1}{1536} (-631+22 x^2+90 x^4) x\tanh  x \lp
+\frac{1}{6144}(-1248+31717 x^2-4632 x^4) \tanh^2 x
+\frac{1}{6144} (-65255+10264 x^2+1296 x^4) x\tanh^3 x \lp
+\frac{1}{2048}(11200-43599 x^2+2174 x^4) \tanh^4 x
-\frac{1}{512}  (-21305+1710 x^2+78 x^4) x\tanh^5 x \lp
+\frac{1}{256} (-5648+7391 x^2-98 x^4) \tanh ^6 x
+\frac{1}{512}  (-27907+864 x^2) x\tanh ^7 x \lp
-\frac{3}{512} (-5080+2151 x^2) \tanh^8 x
+\frac{1479}{64} x \tanh^9 x
-\frac{825}{64} \tanh ^{10} x.
\ea

\paragraph{Remark}

As in (\ref{1.6}), all cumulants can be extracted from $\langle e^{-\Delta\hat L}\rangle$ by differentiation
with respect to $\Delta$, which is the same as in (\ref{5.20}), up to sign. Hence, we can compare our results with what can be 
extracted from (\ref{1.9}). In the case of the variance, we can match only the classical term $\frac{1}{\l}\tanh^{2}(Jt)$  from 
\be
\frac{1}{\l^{2}}(\langle \hat L^{2}\rangle-\langle \hat L\rangle^{2}) = 
\frac{1}{\l^{2}}\frac{d^{2}}{d\Delta^{2}}\log\langle e^{-\Delta \hat L}\rangle|_{\Delta=0} = \mc A_{2}\,\frac{1}{\l}+O(\l^{0}).
\ee
Using (\ref{1.12}) with $\tau=it$ agrees with $K_{0}^{(2)}(Jt)$ (with $J=1/2$). Our loop corrections go beyond the approximation in (\ref{1.9}).

In the case of the third-order cumulant, we cannot use (\ref{1.6}) to get even the classical term
because we would need the $O(\Delta^{3})$ contribution in (\ref{1.9}), which is not known. Hence, in this case, our classical and 1-loop
results are again novel.

\subsubsection{Comparison with numerics}

\begin{figure}[htbp]
    \centering

    \begin{subfigure}{0.4\linewidth}
        \centering
        \includegraphics[width=\linewidth]{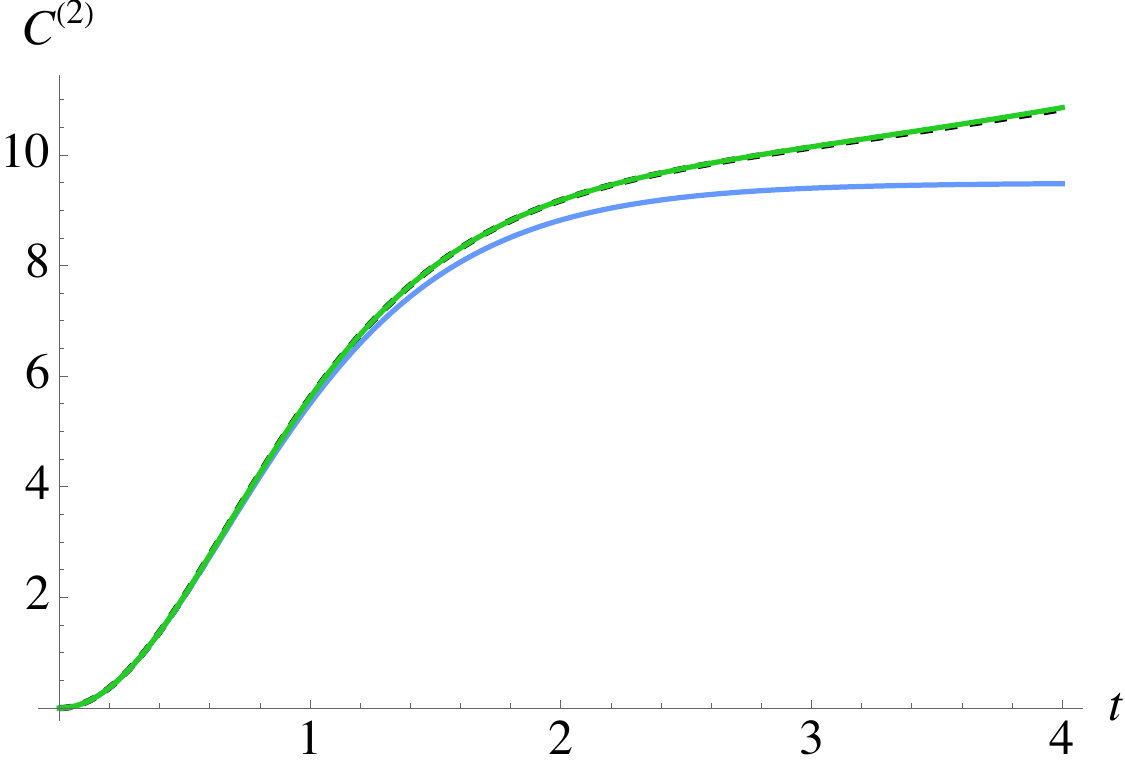}
        %\caption{$\l=2/3$}
%        \label{fig:sub1}
    \end{subfigure}
    \qquad
    \begin{subfigure}{0.4\linewidth}
        \centering
        \includegraphics[width=\linewidth]{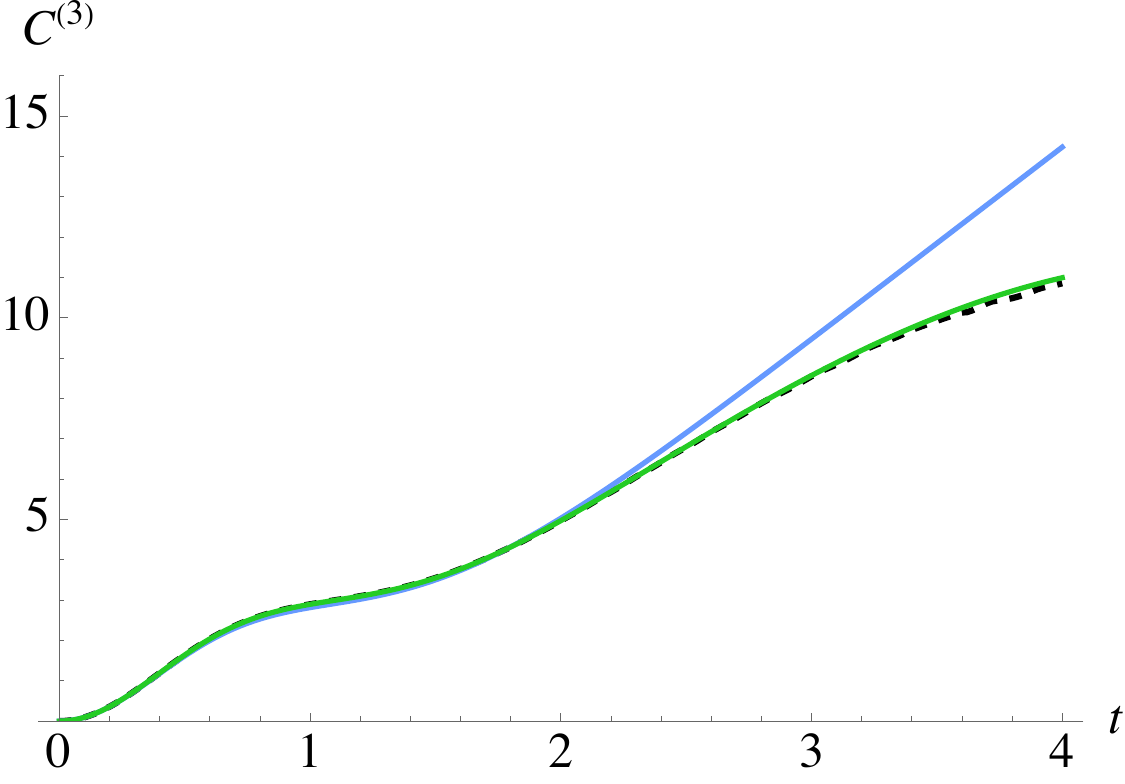}
        %\caption{$\l=1$}
%        \label{fig:sub2}
    \end{subfigure}

    \caption{
    Comparison between the numerical evaluation of the Krylov variance and third-order cumulant at $J=1$ and $q = 1-1/10$.
    In both panels, the black dashed line is the result of integration of (\ref{2.21}) with a truncated Krylov basis with $10^{3}$ states. 
    The blue line is the leading-order $O(1/\l)$ term, while the green line includes the NLO correction at $O(\l^{0})$. 
    }
    \label{fig:cum}
\end{figure}

To test our expansions, we illustrate in this section the numerical evaluation of the Krylov variance and third-order cumulant.
We choose  $J=1$ (which just sets the time scale) and $q = 1-1/10$, corresponding to  $\l \simeq 0.1$, which is not too close to zero so that the corrections are
visible.
The numerical data are computed by solving the set of differential equations (\ref{2.21}) with a cutoff on the maximal Krylov index 
set to $10^{3}$. Larger values give differences that are not appreciable in the explored range of values of $t$. Results are shown in Fig.~\ref{fig:cum}.
They show that the computed next-to-leading contribution is enough to reproduce  the numerical data quite well.

\section{Asymptotic behaviour at small and large time}
\la{sec:asym}

It is of some interest to examine the small and large time behaviour of the Krylov complexity in order to compare with the estimate (\ref{2.24})
obtained in \cite{Rabinovici:2023yex}. 

\begin{figure}[htbp]
    \centering
    \begin{subfigure}{0.4\linewidth}
    \centering
    \includegraphics[width=\linewidth]{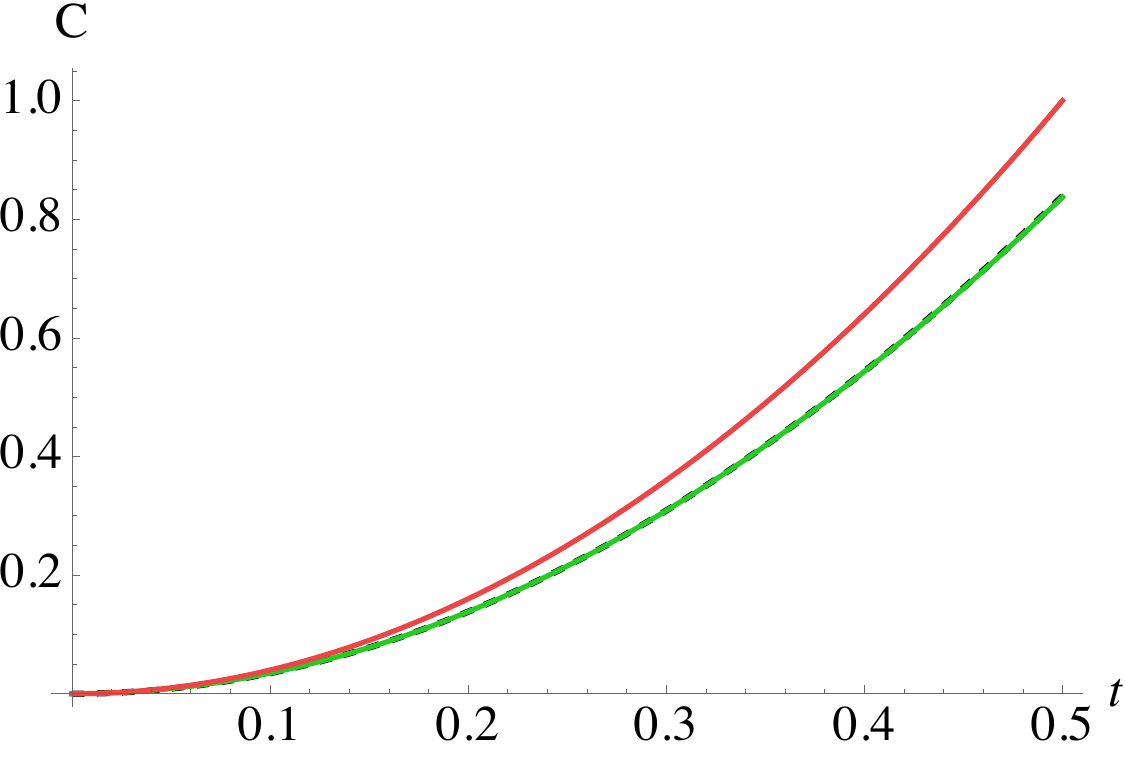}
    \end{subfigure}
    \qquad
    \begin{subfigure}{0.4\linewidth}
    \centering
    \includegraphics[width=\linewidth]{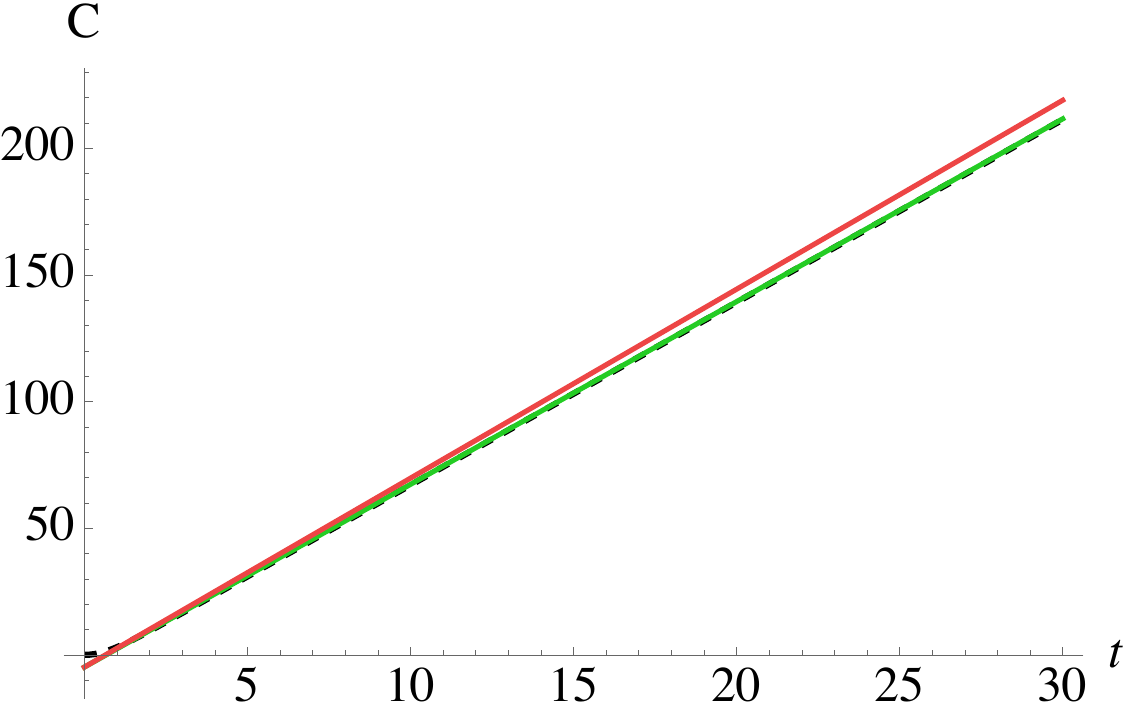}
    \end{subfigure}
    \caption{
    \textbf{Left panel:} small time behaviour of the Krylov complexity at $J=1$ and $q = 1-1/4$.
    The black dashed line is the result of integration of (\ref{2.21}) with a truncated Krylov basis with $10^{3}$ states. 
    The red line is the quadratic function $\frac{1}{1-q}(Jt)^{2}$. The green line is our expansion in (\ref{7.11}).
    \textbf{Right panel:} large time behaviour of the Krylov complexity at $J=1$ and $q = 1-1/4$.
    The black dashed line is the result of integration of (\ref{2.21}) with a truncated Krylov basis with $5\times 10^{3}$ states. 
    The red line is the linear function $\frac{2}{\sqrt{\l(1-q)}}Jt-\frac{2\log 2}{\l}$. The green line is formula (\ref{7.17}) with the addition of 
    the leading time-independent term $-\frac{2\log 2}{\l}$.
    }
    \label{fig:K}
\end{figure}

\paragraph{Small time} At small $t$, we can use the general expression (\ref{5.26}) for $p=1,2,3,\dots$. In the specific case of the Krylov complexity ($p=1$) and
variance $p=2$, this gives 
\be
\la{7.11}
C(t) = \frac{(Jt)^{2}}{\l}-\frac{1-q}{6}\frac{(Jt)^{4}}{\l^{2}}+O(t^{6}), \qquad C^{(2)}(t) = \frac{(Jt)^{2}}{\l}-\frac{2}{3}(1-q)\frac{(Jt)^{4}}{\l^{2}}+O(t^{6}).
\ee
In the case of $C(t)$, apart from the $t^{4}$ correction, we remark that the exact $t^{2}$ coefficient is not equal to $\frac{1}{1-q}$ in the estimate (\ref{2.24}), although they are 
equal in the limit $\l\to 0$.
We believe the reason is that (\ref{2.24}) is obtained from the small and large $\ell$ behaviour of the Lanczos coefficients $b_{\ell}$ in (\ref{2.22}). However, this is sensitive to the discrete nature of $\ell$.
For instance, at small time, the analysis in \cite{Rabinovici:2023yex} is based on approximations like $q^{\ell} = e^{-\ell\,\l} \simeq 1$. 
This is certainly valid when $\ell$ is finite and $\l\to 0$. But if $\l$ is finite, requiring $\ell$ small is unclear due to discreteness,
as is clear from considering the minimal value $\ell=1$. Notice that the small time estimate in (\ref{2.24}) was tested numerically 
in \cite{Rabinovici:2023yex} at moderate times and $q$ close to 1. We analyze numerically the complexity
at smaller times and $q$ not so close to 1. This is shown in the left panel of Fig.~\ref{fig:K} 
where one sees that (\ref{7.11})  reproduces the numerical data very well. 

\paragraph{Large time}
Finding the behaviour at large time is non-trivial, but from the 5-loop expression of the Krylov complexity, 
we see that -- up to exponentially small corrections \ie setting $\tanh(Jt)\simeq 1$ --
one has (notation for suppressed corrections is schematical)
\ba
\la{8.2}
C(t) &= A_{0}(\l)+A_{1}(\l)\, Jt + O(e^{-Jt}), 
\ea
with
\bea
\la{8.3}
A_{1}(\l) &= \frac{2}{\l}+\frac{1}{4}+\frac{7\l}{192}+\frac{5\l^{2}}{1536}-\frac{\l^{3}}{81920}-\frac{101\l^{4}}{5898240}+\cdots, \\
A_{0}(\l) &= -\frac{2\log 2}{\l}-\frac{1}{4}-\frac{7\l}{96}-\frac{\l^{2}}{64}-\frac{211\l^{3}}{46080}-\frac{7\l^{4}}{3072}+\cdots.
\eea
All higher powers of $t$ cancel and the leading large $t$ behaviour is linear in $t$ with $\l$-dependent coefficients.  \footnote{
This linear behaviour is consistent with (\ref{6.15}) after $\mu\to it$ where the r.h.s. vanishes exponentially. However, the determination of the slope and intercept
coefficients as functions of $\l$ is non-trivial because they are sensitive to the boundary conditions of the operator Liouville equation, which are set at $t=0$.}
The expression (\ref{8.3}) is valid within our small $\l$ expansion, \ie for $\l\to 0$. A suggestive conjecture for the generating function of the $\l$-series
of slope $A_{1}(\l)$ is 
\be
\la{7.17}
A_{1}^{\rm pert}(\l) = \frac{2}{\l}e^{\l/8}\sqrt\frac{\l/4}{\tanh(\l/4)},
\ee
where the label ``pert'' reminds that this reproduces the small $\l$ expansion of the true slope.
These expressions should be compared with (\ref{2.24}); 
agreement holds only as $\l\to 0$,  which is expected since the large-time estimate in (\ref{2.24}) captures only the leading ballistic velocity of the wave packet along the Krylov chain, 
missing the corrections from the spreading tail of the distribution, as noted in \cite{Rabinovici:2023yex}.
The numerical analysis of (\ref{7.17}) is presented in the right 
panel of Fig.~(\ref{fig:K}) for $q=1-1/4$, \ie the moderate value $\l=0.29$.

\begin{figure}[htbp]
    \centering
    \begin{subfigure}{0.4\linewidth}
    \centering
    \includegraphics[width=\linewidth]{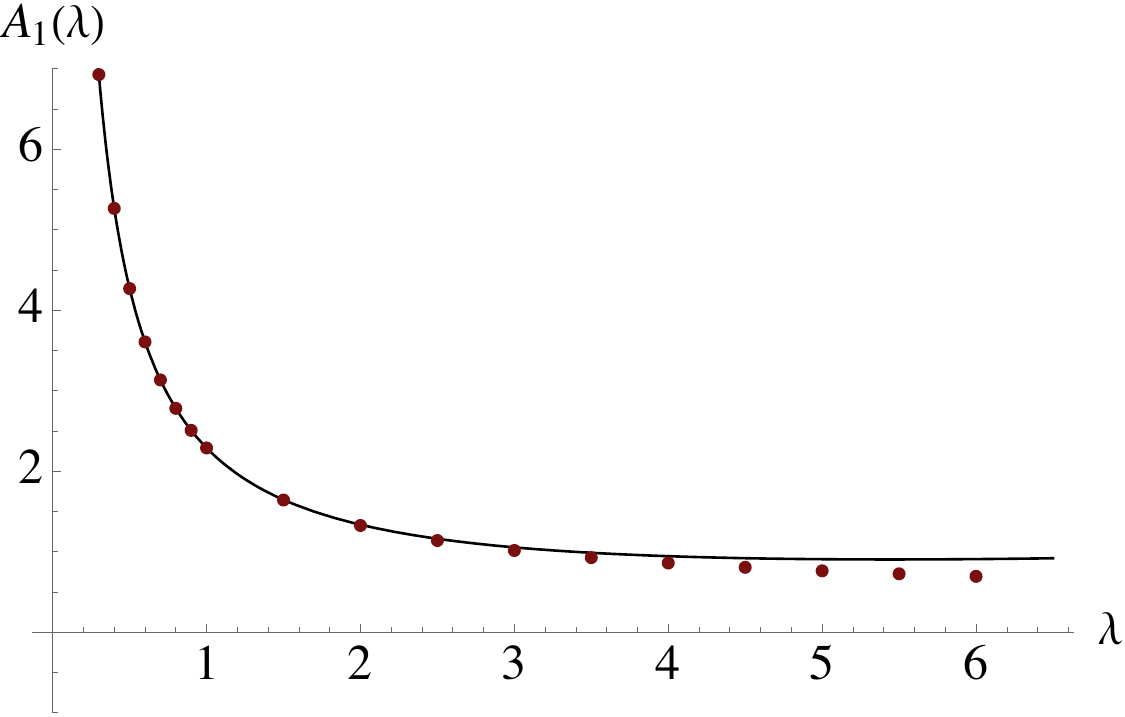}
    \end{subfigure}
    \qquad
    \begin{subfigure}{0.4\linewidth}
    \centering
    \includegraphics[width=\linewidth]{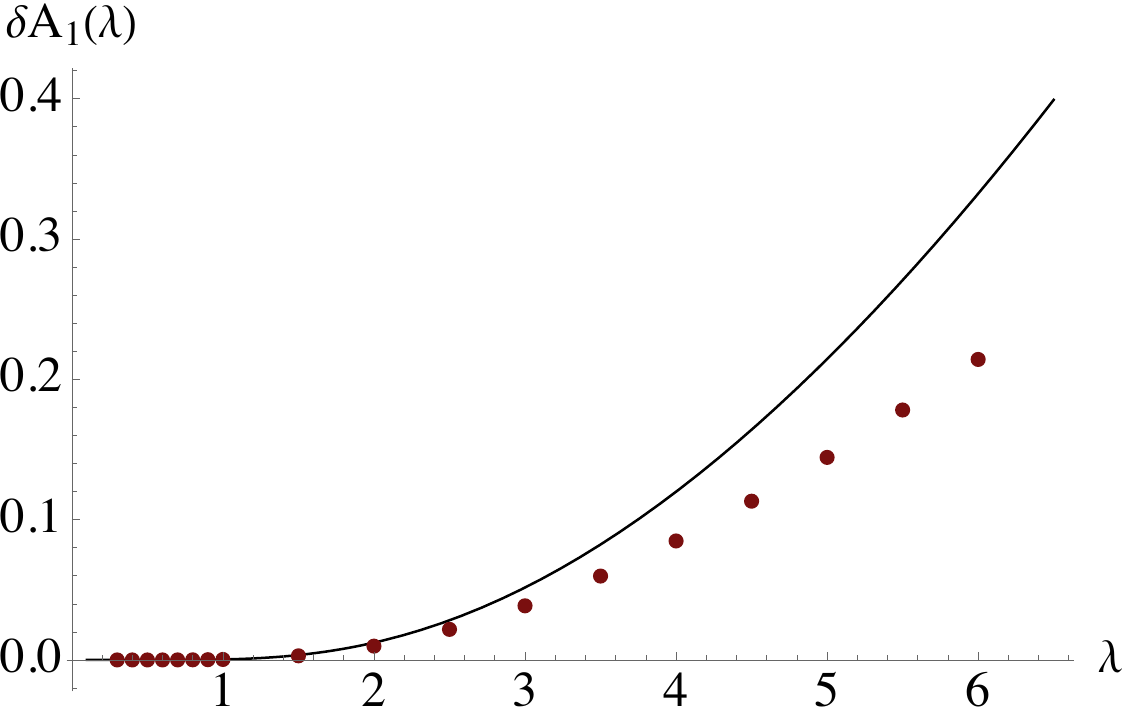}
    \end{subfigure}
    \caption{
    Comparison between the numerically estimated slope $A_{1}(\l)$ and the generating function $A_{1}^{\rm pert}(\l)$ in (\ref{7.17}). The numerics
    are performed by using a Krylov space dimension equal to $6000$ and studying $C'(t)$, taking  the plateau value of $C'(t)$
     at intermediate times as an estimate.
    \textbf{Left panel:} The solid thin line is (\ref{7.17}), while red dots are numerical data.    
    \textbf{Right panel:} Red dots are the difference between (\ref{7.17}) and numerical data. The thin solid black line is the leading non-perturbative correction in (\ref{815}).
        }
    \label{fig:NP}
\end{figure}

To gain more information, we compute numerically the time derivative of the Krylov complexity $C'(t)$ using a Krylov basis with $6000$ states and 
taking $t$ large enough to observe a plateau (or a maximum). This value is our numerical estimate of the slope $A_{1}(\l)$. We then compare
it with the generating function in (\ref{7.17}). We find that it is rather accurate up to $\l\sim 1$, where relative error is below the permille level, 
while for larger $\l$ a systematic deviation is observed,
which is not a finite-size artifact, as confirmed by varying the Krylov space dimension.
It appears to be a non-perturbative correction vanishing exponentially fast for $\l\to 0$, and 
reminiscent of non-perturbative corrections to the DSSYK partition function \cite{Okuyama:2025fhi}.
This analysis is shown in Fig.~\ref{fig:NP}, see details in the caption.

Actually, we can determine an exact expression for the slope $A_{1}$ by stationary phase approximation in large time limit, as shown in Appendix~\ref{app:slope}. 
This gives the exact 
expression 
\be
\la{85}
A_{1}(\l) = (q;q)_{\infty}\int_{0}^{\pi}\frac{d\theta}{2\pi}\frac{2J\sin\theta}{\sqrt{\l(1-q)}}|(e^{2i\theta};q)_{\infty}|^{2},
\ee
whose numerical evaluation reproduces all data points.
Expanding in small $q$, or large $\l$,  gives
\be
A_{1}(\l) = \frac{16}{3\pi\sql}\bigg(1+\frac{7}{10}e^{-\l}+\frac{19}{40}e^{-2\l}+\cdots\bigg).
\ee
The expansion in $\l\to 0$ limit is a little harder. 
The first (asymptotic) expansion we need is \cite{Okuyama:2023bch}
\be
\la{U1}
(z; q)_{\infty} = \exp\bigg[-\sum_{n=1}^{\infty}\frac{z^{n}}{n(1-q^{n})}\bigg] \sim \exp\bigg[-\sum_{g=0}^{\infty}\frac{B_{2g}}{(2g)!}\text{Li}_{2-2g}(z)\, \l^{2g-1}+\frac{1}{2}\log(1-z)\bigg].
\ee
Here it is useful to use 
\be
\text{Li}_{2-2g}(e^{2i\theta})+\text{Li}_{2-2g}(e^{-2i\theta}) = \begin{cases}
2(\theta-\frac{\pi}{2})^{2}-\frac{\pi^{2}}{6}, & g=0, \\
-1, & g=1, \\
0, & g\ge 2.
\end{cases}
\ee
This implies 
\ba
\la{89}
|(e^{2i\theta};q)_{\infty}|^{2} &\sim \exp\bigg[-\sum_{g=0}^{\infty}\frac{B_{2g}}{(2g)!}(\text{Li}_{2-2g}(e^{2i\theta})+\text{Li}_{2-2g}(e^{-2i\theta})) \l^{2g-1}
+\frac{1}{2}\log(1-e^{2i\theta})+\frac{1}{2}\log(1-e^{-2i\theta})\bigg]\lp
= 2\sin\theta \exp\bigg\{-\frac{1}{\l}\bigg[2\bigg(\theta-\frac{\pi}{2}\bigg)^{2}-\frac{\pi^{2}}{6}\bigg]+\frac{\l}{12}\bigg\}.
\ea
Corrections are non-perturbative and read at leading order, see Appendix~\ref{app:expansion},
\ba
\la{810}
|(e^{2i\theta};q)_{\infty}|^{2} &
= 2\sin\theta \exp\bigg\{-\frac{1}{\l}\bigg[2\bigg(\theta-\frac{\pi}{2}\bigg)^{2}-\frac{\pi^{2}}{6}\bigg]+\frac{\l}{12}\bigg\}\bigg[1-2e^{-\frac{2\pi^{2}}{\l}}\cosh\frac{4\pi(\theta-\pi/2)}{\l}+\cdots\bigg].
\ea
The next piece we need is -- using modular transformation of Dedekind function -- 
\be
(q;q)_{\infty} = q^{-1/24}\eta(q) = \sqrt\frac{2\pi}{\l}\exp\bigg(-\frac{\pi^{2}}{6\l}+\frac{\l}{24}\bigg)\,(\wt q; \wt q)_{\infty}, \qquad \wt q = e^{-4\pi^{2}/\l}.
\ee
The last factor is also a (subleading) non-perturbative correction
\be
(\wt q; \wt q)_{\infty} = 1-e^{-4\pi^{2}/\l}-e^{-8\pi^{2}/\l}+e^{-20\pi^{2}/\l}+\cdots.
\ee
This gives
\ba
A_{1}(\l) &=  \frac{2J}{\l}e^{\l/8}\sqrt\frac{2}{\pi} \frac{(\wt q; \wt q)_{\infty}}{\sqrt{1-q}}\int_{0}^{\pi}d\theta\sin^{2}\theta\, \exp\bigg[-\frac{2}{\l}\bigg(\theta-\frac{\pi}{2}\bigg)^{2}\bigg]
\bigg[1-2e^{-\frac{2\pi^{2}}{\l}}\cosh\frac{4\pi(\theta-\pi/2)}{\l}+\cdots\bigg].
\ea
The perturbative part is then computed by dropping non-perturbative corrections and extending the integration range to $(-\infty,\infty)$. We get 
\ba
\la{814}
A_{1}^{\rm pert}(\l) &=  \frac{2J}{\l}e^{\l/8}\sqrt\frac{2}{\pi} \frac{1}{\sqrt{1-q}}\int_{-\infty}^{\infty}d\theta\cos^{2}\theta\, \exp\bigg(-\frac{2\theta^{2}}{\l}\bigg) = 
\frac{2}{\l}e^{\l/8}\sqrt{\frac{\l/4}{\tanh(\l/4)}},
\ea
that prove the conjecture (\ref{7.17}). Computing the exact integral in $\theta$, the leading non-perturbative correction is 
\be
\la{815}
A_{1}=A_{1}^{\rm pert}-\frac{2\sqrt 2}{\pi^{7/2}}\l^{3/2}e^{-\frac{\pi^{2}}{2\l}}+\cdots.
\ee
For completeness, we also present the large time expansion of the Krylov variance and third-order cumulant.
They read -- again up
to exponentially suppressed terms -- 
\be
C^{(p)}(t) = \sum_{n=0}^{p}A_{n}^{(p)}(\l)(Jt)^{n},
\ee
where, for $p=2,3$ we have the small $\l$ expansions
\bea
\la{x817}
A_{0}^{(2)}(\l) &= \frac{1}{\l}+\frac{3}{8}+\frac{\l}{16}-\frac{\l^{3}}{192}+\cdots, \qquad
A_{1}^{(2)}(\l) = -\frac{1}{4}+\frac{\l}{32}+\frac{49\l^{2}}{1536}+\frac{55\l^{3}}{4096}+\cdots, \\
A_{2}^{(2)}(\l) &= \frac{1}{8}-\frac{\l}{32}-\frac{13\l^{2}}{1536}-\frac{\l^{3}}{6144}+\cdots,
\eea
and
\bea
\la{x818}
A_{0}^{(3)}(\l) &= -\frac{1}{2\l}+\frac{25\l}{192}+\frac{5\l^{2}}{64}+\cdots, \qquad
A_{1}^{(3)}(\l) = \frac{1}{2\l}-\frac{5}{16}-\frac{83\l}{768}+\frac{29\l^{2}}{6144}+\cdots, \\
A_{2}^{(3)}(\l) &= \frac{3}{8}-\frac{3\l}{32}-\frac{17\l^{2}}{256}+\cdots, \qquad
A_{3}^{(3)}(\l) = -\frac{1}{8}+\frac{9\l}{128}+\frac{\l^{2}}{256}+\cdots.
\eea
The same calculation as in (\ref{814}) \footnote{In the notation of (\ref{B16}), we use  
$\lim_{t\to\infty}\frac{C^{(p)}(t)}{t^{p}} =  \int_0^\pi v(\theta)^{p}\,\psi_0(\theta)^2\,d\theta$.
}
gives the exact resummation of the leading coefficients at large time
\bea
\la{res-cum}
A_{2}^{(2), {\rm pert}}(\l) &= -\frac{1}{\l}\frac{1-2 e^{3 \lambda /4}+e^{\lambda }}{(1+e^{\lambda /4}) \
(1+e^{\lambda /2})}, \\
A_{3}^{(3), {\rm pert}}(\l) &=\frac{1}{\l^{3/2}} \frac{e^{-3 \lambda /8} (-1+e^{\lambda /4})^2 (-1-3 e^{\lambda /4}-2 \
e^{\lambda /2}+2 e^{\lambda })}{(1+e^{\lambda /4})   
\sqrt{e^{\lambda }-1}},
\eea
with expansions
\ba
A_{2}^{(2), {\rm pert}}(\l) &= \frac{1}{8}-\frac{\lambda }{32}-\frac{13 \lambda 
^2}{1536}-\frac{\lambda ^3}{6144}+\frac{31 \lambda 
^4}{245760}-\frac{\lambda ^5}{2949120}-\frac{2201 \lambda 
^6}{660602880}+\cdots, \\
A_{3}^{(3), {\rm pert}}(\l) &= -\frac{1}{8}+\frac{9 \lambda }{128}+\frac{\lambda ^2}{256}+\frac{9 
\lambda ^3}{16384}-\frac{59 \lambda ^4}{3932160}+\frac{27 \lambda 
^5}{4194304}+\frac{11651 \lambda ^6}{15854469120}+\cdots,
\ea
in agreement with (\ref{x817}, \ref{x818}).

\paragraph{Acknowledgements}

We thank K. Okuyama and K. Suzuki for clarifications. We also thank P. Nandy, R. N. Das,  S. E. Aguilar-Gutierrez,  J. Papalini, and W. M\"uck
for useful comments related to this work.
MB is supported by the INFN grant GAST. EA is supported by the MUR project GINEVRA, prot.
2022BZYBWM.

\appendix

\section{Notation for $q$-functions}
\la{app:q-functions}

We use
\ba
& (z;q)_{n} = \prod_{k=0}^{n-1}(1-z q^{k}), \qquad
(z_{1}, \dots, z_{k}; q)_{n} = \prod_{i=1}^{k}(z_{i}; q)_{n}, \\
& (e^{\pm i\theta}; q)_{n}\equiv (e^{ i\theta}; q)_{n}(e^{- i\theta}; q)_{n} = |(e^{i\theta}; q)|^{2},
\ea
and the $q$-Hermite polynomials 
\be
\la{A.3}
H_{n}(\cos\theta|q) \equiv \sum_{k=0}^{n}\frac{(q;q)_{n}}{(q;q)_{n-k}(q;q)_{k}}e^{i(n-2k)\theta}.
\ee

\section{Large time slope of Krylov complexity for models with $b_{n}\to b_{\infty}$}
\la{app:slope}

Let $H$ be a Hamiltonian in Krylov form 
\be
\la{tridiagonal}
H|n\rangle = b_{n+1}|n+1\rangle + b_n|n-1\rangle, \qquad b_0\equiv 0 ,
\ee
with real Lanczos coefficients $b_n\geq0$. To recall, the Krylov complexity for the initial state $\ket{0}$ is
\be
\la{CK-def}
C(t) = \sum_n n\,|\phi_n(t)|^2 ,
\ee
where the time-evolved state $|\phi(t)\rangle = e^{-i Ht}|0\rangle$ has Krylov-basis components $\phi_n(t)=\langle n|\phi(t)\rangle$.
We assume throughout that $b_n\to b_\infty$ as $n\to\infty$, with $|b_n-b_\infty|\to 0$ sufficiently rapidly.
Then (\ref{tridiagonal}) has a continuous spectrum filling a band
\be
\la{dispersion}
E(\theta) = 2b_\infty\cos\theta, \qquad \theta\in(0,\pi),
\ee
with (real, orthonormal) generalized eigenfunctions $\psi_n(\theta)\equiv\langle n|\theta\rangle$ satisfying
\be
\int_0^\pi d\theta\, \psi_n(\theta)\psi_m(\theta) = \delta_{nm}, \qquad \sum_n \psi_n(\theta)\psi_n(\theta') = \delta(\theta-\theta') .
\ee
In this basis,
\be
\la{phin}
\phi_n(t) = \int_0^\pi d\theta\; e^{-i E(\theta)t}\,\psi_n(\theta)\,\psi_0(\theta) ,
\ee
and we want to prove that 
\be
\la{main-result}
\lim_{t\to\infty}\frac{C(t)}{t} = \int_0^\pi d\theta\; v(\theta)\,\psi_0(\theta)^2, \qquad v(\theta) \equiv \big|E'(\theta)\big| = 2b_\infty\sin\theta ,
\ee
\ie the late-time growth rate is the group velocity $v(\theta)$, averaged over the \emph{exact} spectral weight $\psi_0(\theta)^2$ of the reference state under the energy eigenbasis -- 
regardless of any other details of the Lanczos sequence.

To this aim, we start by examining the large-$n$ behavior of $\psi_n(\theta)$ at fixed $\theta$. For $n$ large enough that $b_n\approx b_\infty$, the recursion
\be
E\,\psi_n(E) = b_{n+1}\psi_{n+1}(E) + b_n\psi_{n-1}(E),
\ee
becomes, to leading order, the free recursion $E\psi_n = b_\infty(\psi_{n+1}+\psi_{n-1})$, whose general solution at $E=2b_\infty\cos\theta$ is a linear combination of $e^{\pm i n\theta}$. 
This is the discrete analogue of a one-dimensional scattering (Jost) problem: (\ref{tridiagonal}) is a half-line ($n\geq0$) hopping problem with a ``defect'' region 
(the deviation of $b_n$ from $b_\infty$ at small/moderate $n$) and a hard wall at $n=-1$ ($b_0=0$). 
Because the chain terminates rather than extending to $n\to-\infty$, there is no transmission channel: the physical solution selected by the $n=-1$ boundary condition is, for large $n$, a pure standing wave,
\be
\la{jost}
\psi_n(\theta) \;\longrightarrow\; \sqrt\frac{2}{\pi}\,\sin\!\big(n\theta+\delta(\theta)\big), \qquad n\to\infty ,
\ee
where the amplitude is fixed from completeness relation. The phase shift $\delta(\theta)$ encodes the entire ``defect'' (\ie 
all of the Lanczos coefficients that differ from $b_\infty$) and is in general a complicated, 
model-dependent function of $\theta$. 

We now insert (\ref{jost}) into (\ref{phin}) and write the sine as a difference of exponentials:
\be
\label{phin-split}
\phi_n(t) \;\approx\; \frac{1}{i\sqrt{2\pi}}\int_0^\pi d\theta\,\psi_0(\theta)\,
\Big[e^{i(n\theta+\delta(\theta)-E(\theta)t)} - e^{-i(n\theta+\delta(\theta)+E(\theta)t)}\Big] .
\ee
We seek the regime $n,t\to\infty$ with $x\equiv n/t$ fixed. The stationary points of the two phases are
\ba
\frac{d}{d\theta}\big[n\theta \mp E(\theta)t\big]=0 \;\Longrightarrow\; n = \pm E'(\theta)\,t.
\ea
Since $E'(\theta)=-2b_\infty\sin\theta<0$ on $(0,\pi)$, the branch with $n,t>0$ is the relevant one, namely \footnote{
The $\delta(\theta)$ term in the phase shifts the saddle by a relative $O(1/t)$ correction, which does not affect the leading large-$t$ asymptotics.
}
\be
\la{saddle}
n = v(\theta)\,t, \qquad v(\theta)\equiv-E'(\theta) = 2b_\infty\sin\theta>0 ,
\ee
where $\theta$ is now a function of $n/t=x$.
Equation (\ref{saddle}) is  the statement that the energy mode $\theta$ propagates at its group velocity $v(\theta)$.
Standard stationary-phase asymptotics applied to the second term of (\ref{phin-split}) give
\be
\phi_n(t) \;\approx\; -\frac{1}{i\sqrt{2\pi}}\,\psi_0(\theta)\,e^{-i\Phi(\theta)}\,e^{i\frac{\pi}{4}\,\mathrm{sgn}(E''(\theta))}\sqrt{\frac{2\pi}{t\,|E''(\theta)|}} ,
\ee
where $\Phi(\theta)=n\theta+E(\theta)t$, so that
\be
|\phi_n(t)|^2 \;\approx\; \psi_0(\theta)^2\,\frac{1}{t\,|E''(\theta)|} .
\ee
Multiplying by $t$ defines the limiting probability density $P(x)$ for $x=n/t$ (so that $\sum_n |\phi_n(t)|^2 \approx \int d n\,|\phi_n(t)|^2 = \int d x\; t\,|\phi_{xt}(t)|^2 \to \int d x\, P(x)$):
\be
P(x) = t\,|\phi_n(t)|^2 = \frac{\psi_0(\theta)^2}{|E''(\theta)|} .
\ee
Changing variables from $x$ to $\theta$ via $x=v(\theta)$, so that $d x = |v'(\theta)|\,d\theta = |E''(\theta)|\,d\theta$, the Jacobian cancels the $1/|E''(\theta)|$ factor exactly, giving
\be
\la{pushforward}
P(x)\,d x \;=\; \psi_0(\theta)^2\,d\theta, \qquad x=v(\theta),
\ee
\ie the limiting distribution of $n/t$ is exactly the pushforward of the exact spectral measure $\psi_0(\theta)^2\,d\theta$ under the group-velocity map $\theta\mapsto v(\theta)$, with \emph{no} dependence on the phase shift $\delta(\theta)$ and no approximation beyond the stationary-phase (large-$t$) limit itself.

By (\ref{CK-def}) and (\ref{logX-exact}), we get 
\be
\la{B16}
\frac{C(t)}{t} = \sum_n \frac{n}{t}\,|\phi_n(t)|^2 \;\xrightarrow{t\to\infty}\; \int x\,P(x)\,d x = \int_0^\pi v(\theta)\,\psi_0(\theta)^2\,d\theta ,
\ee
which is (\ref{main-result}). 

\paragraph{Consistency check: the flat Lanczos sequence}

As a check, take $b_n\equiv b$ for all $n$ (so $b_\infty=b$), for which $E(\theta)=2b\cos\theta$ and the exact eigenfunctions are $\psi_n(\theta)=\sqrt{2/\pi}\sin\big((n+1)\theta\big)$, hence $\psi_0(\theta)^2=(2/\pi)\sin^2\theta$. Then, we get 
\be
\la{flat-check}
\lim_{t\to\infty}\frac{C(t)}{t} = \int_0^\pi \frac{2}{\pi}\sin^2\theta\cdot 2b\sin\theta\,d\theta = \frac{4b}{\pi}\int_0^\pi \sin^3\theta\,d\theta = \frac{4b}{\pi}\cdot\frac{4}{3} = \frac{16b}{3\pi} .
\ee
This matches the known exact asymptotic, obtained independently from the closed-form wavefunction $\phi_n(t)=i^n\frac{n+1}{bt}J_{n+1}(2bt)$ (a Bessel function) 
and its large-argument expansion, $C(t)\to \tfrac{16}{3\pi}bt + O(1)$ \cite{Barbon:2019wsy,Rabinovici:2023yex}, see also \cite{Muck:2022xfc}. 
In particular, (\ref{flat-check}) correctly differs from the naive ``ballistic front'' estimate $v_{\max}t=2bt$ (the maximum of $v(\theta)$, attained at $\theta=\pi/2$): the true late-time slope is the $v(\theta)$-average weighted by $\psi_0(\theta)^2$, not the maximum group velocity.

\paragraph{Specialization to DSSYK}

We now apply (\ref{B16}) to the genuinely $n$-dependent Lanczos sequence of DSSYK. For the Lanczos coefficients in (\ref{2.11}) 
$b_n - b_\infty = O(q^n)$ decays geometrically and
\be
b_\infty = \frac{J}{\sqrt{\lambda(1-q)}} , \qquad v(\theta) = 2b_\infty\sin\theta = \frac{2J\sin\theta}{\sqrt{\lambda(1-q)}} .
\ee
The exact spectral weight of the reference (zero-chord) state is known in closed form,
\be
\psi_0(\theta)^2 = \frac{(q;q)_\infty}{2\pi}\,\big|(e^{2i\theta};q)_\infty\big|^2 .
\ee
We then get 
\be
\lim_{t\to\infty}\frac{C(t)}{t} = \frac{J\,(q;q)_\infty}{\pi\sqrt{\lambda(1-q)}}\int_0^\pi d\theta\;\sin\theta\,\big|(e^{2i\theta};q)_\infty\big|^2\;,
\ee
which is (\ref{85}).

\section{Non-perturbative corrections to $|(e^{\pm 2i\theta},q)_{\infty}|^{2}$}
\la{app:expansion}

Let us denote $X(\theta)\equiv\big|(e^{2i\theta};q)_\infty\big|^2$ and find its small $\l$ expansion, including non-perturbative corrections.
As a first step, we write 
\be
\log X(\theta) = \log(e^{2i\theta};q)_\infty + \log(e^{-2i\theta};q)_\infty = 2\,\mathrm{Re}\big[\log(z;q)_\infty\big]_{z=e^{2i\theta}} ,
\label{logX-Re}
\ee
The Mittag-Leffler expansion of $1/(1-e^{-x})$,
\be
\frac{1}{1-e^{-x}} = \frac1x+\frac12+\sum_{j=1}^\infty\frac{2x}{x^2+4\pi^2j^2} ,
\ee
substituted into $\log(z;q)_\infty=-\sum_{n\geq1}\frac{z^n}{n}\frac{1}{1-q^n}$ at $x=n\lambda$, $z=e^{i\varphi}$, gives the exact rewriting
\be
\log(z;q)_\infty = -\frac1\lambda\mathrm{Li}_2(z) + \frac12\log(1-z) - \frac2\lambda\sum_{j=1}^\infty S(\gamma_j,\varphi) , \qquad S(\gamma,\varphi)\equiv\sum_{n=1}^\infty\frac{e^{i n\varphi}}{n^2+\gamma^2},\ \ \gamma_j\equiv\frac{2\pi j}{\lambda} .
\ee
We now use the sum 
\be
\la{green}
\sum_{n=-\infty}^\infty \frac{e^{i n\varphi}}{n^2+\gamma^2} = \frac\pi\gamma\,\frac{\cosh\big(\gamma(\pi-d(\varphi))\big)}{\sinh(\pi\gamma)} , \qquad d(\varphi)\equiv\min(\varphi,2\pi-\varphi)\in(0,\pi) ,
\ee
valid for $\varphi\in(0,2\pi)$. Since the left side equals $1/\gamma^2+2\,\mathrm{Re}[S(\gamma,\varphi)]$, (\ref{green}) 
determines $\mathrm{Re}[S(\gamma,\varphi)]$ in closed form, exactly, for every $\gamma$.
Combining (\ref{logX-Re}--\ref{green}) with $\varphi=2\theta$, $d(2\theta)=2\min(\theta,\pi-\theta)$, 
and $\sum_j\gamma_j^{-2}=\tfrac{\lambda^2}{4\pi^2}\cdot\tfrac{\pi^2}{6}=\tfrac{\lambda^2}{24}$ 
gives
\be
\la{logX-exact}
\log X(\theta) = -\frac2\lambda\,\mathrm{Re}\big[\mathrm{Li}_2(e^{2i\theta})\big] + \log(2\sin\theta) + \frac\lambda{12} \;-\; \frac2\lambda\sum_{j=1}^\infty F(\gamma_j,2\theta) , \qquad F(\gamma,\varphi)\equiv\frac\pi\gamma\,\frac{\cosh(\gamma(\pi-d))}{\sinh(\pi\gamma)} .
\ee
Using $\mathrm{Re}[\mathrm{Li}_2(e^{2i\theta})]=2(\theta-\pi/2)^2-\pi^2/6$, the first three terms of (\ref{logX-exact}) are exactly $\log X_{\rm pert}(\theta)$, 
\cf (\ref{89}). For the non-perturbative part,
writing $\cosh(\gamma A)=\tfrac12(e^{\gamma A}+e^{-\gamma A})$, $\sinh(\gamma B)=\tfrac12(e^{\gamma B}-e^{-\gamma B})$ and expanding for large $\gamma$,
\be
F(\gamma,\varphi) = \frac\pi\gamma\Big[e^{-\gamma d}+e^{-\gamma(2\pi-d)}+e^{-\gamma(d+2\pi)}+\cdots\Big] ,
\ee
a purely exponentially small series with no power-law-in-$1/\gamma$ piece. Keeping only the $j=1$, two-leading-exponential terms in 
(\ref{logX-exact}) (using $\pi/\gamma_1=\lambda/2$ and $d(2\theta)=2\min(\theta,\pi-\theta)$),
\be
\log X(\theta) - \log X_{\rm pert}(\theta) \;=\; -e^{-4\pi\theta/\lambda} - e^{-4\pi(\pi-\theta)/\lambda} \;+\; O\big(e^{-8\pi\min(\theta,\pi-\theta)/\lambda}\big) ,
\ee
which gives (\ref{810}). 
Note that the expansion is valid for $\theta\in(0,\pi)$, excluding the 
endpoints $\theta=0,\pi$; since $X(\theta)\sim 2\sin\theta\to0$ there, 
\cf (\ref{logX-exact}), these endpoints do not contribute to the slope 
integral (\ref{85}), as also confirmed numerically.

\bibliography{Krylov-Biblio}
\bibliographystyle{JHEP-v2.9}
\end{document}